\documentclass[letterpaper,twocolumn,10pt]{article}
\usepackage{usenix2019_v3}
\usepackage{amsmath}
\usepackage{amssymb}
\usepackage{graphicx}
\usepackage{booktabs}
\graphicspath{{figures/}}

\begin{document}
\raggedbottom
\date{}
\title{\Large \bf Behind a Simple Read: Understanding Work and Waiting in the Linux I/O Stack}
\author{
{\rm Yang Shen, Wenzhe Zhang\textsuperscript{*}, Yong Dong, Min Xie,}\\
{\rm Wei Zhang, Huijun Wu, and Zhenwei Wu}\\
College of Computer Science and Technology\\
National University of Defense Technology\\
Changsha 410073, China\\
{\small\texttt{\{shenyang\_23,zhangwenzhe,yongdong,xiemin\}@nudt.edu.cn}}\\
{\small\texttt{\{weizhang,wuhuijun,zhenweiwu\}@nudt.edu.cn}}\\
{\small\textsuperscript{*}Corresponding author}
}
\maketitle

\begin{abstract}
A simple \texttt{read} interface provides uniform semantics but unpredictable
performance. Using synchronous large-buffer Linux reads, we decompose the
buffered-read path into work volume, processing time, and critical-path
exposure. Reducing local cost or accelerating request submission need not
shorten a complete read. Linux uses large folios to reduce metadata work and
I/O--copy overlap to hide most cold-read copying; an experimental kernel
prototype shows that more aggressive overlap and parallelism can shorten the
host path when these mechanisms fail.

Device waiting dominates SSD cold reads. We abstract
first-completion wait $F$ and subsequent completion capacity $B_{CQ}$ from
request-batch timelines. Real-SSD experiments characterize their variation
with request and batch size, while MQSim traces explain their internal origins.
A wait chain along the folio--bio--request mapping connects device
completion to upper-layer waiting. Under strict batching, the median absolute
cross-layer prediction error is at most 7.43\%.

Finally, we apply the completion model and wait chain to request supply, Linux
readahead, data layout, and waiting policies. Cases cover gains, no gains, and
reversals, comparing predictions with interventions to close the loop from
observation and modeling to understanding and control and provide a measurable
basis for layered I/O decisions.

\end{abstract}

\section{Introduction}
\label{sec:introduction}

Modern I/O systems meet application performance demands through an expanding
set of software mechanisms and increasingly complex hardware organizations.
On the software side, asynchronous I/O, readahead, cache management, and
multiqueue submission span applications and runtimes, the operating-system
page cache, filesystems, the block layer, and drivers; policies at these
layers interact with one another~\cite{RocksDBBlockCache,RocksDBDirectIO}.
On the hardware side, storage backends continue to diversify, while SSD
caches, parallel resources, garbage collection, wear leveling, and designs
such as computational storage and processing in memory make host-observed
performance harder to predict~\cite{SNIAComputationalStorage,UPMEMTechnology}.
These mechanisms expand system capability, but they also make the process
from expressing a read demand to returning its data more complex, with
interdependent stages throughout the path.

This complexity is not confined to any individual module. Access advice from
an application can change how the kernel organizes requests, the granularity
of cached objects affects management costs, and the software submission pace
changes the work received by the device. These cross-layer effects make
performance bottlenecks difficult to locate and optimization outcomes
difficult to anticipate. In practice, optimization therefore often relies on
rules of thumb such as ``random access should disable readahead'' or ``greater
parallelism improves throughput.'' Yet their effects depend on the workload,
resource state, and execution path; when those conditions change, the same
operation may provide no benefit or even degrade performance. The issue is
not whether experience is valuable, but that the conditions under which it
holds are often left implicit, and an improvement in local behavior does not
necessarily translate into a faster complete read.

We use synchronous, large-buffer Linux buffered reads as an observation
window. This path is triggered by a simple \texttt{read}-like interface, yet
spans the software stack and connects host-side work with device-side waiting;
the analytical methods in this paper are not limited to this interface. Such
encapsulation frees applications from explicitly orchestrating lower-level
operations, but hiding execution details does not make their performance
effects disappear. A simple read interface provides uniform functional
semantics, but not equally simple or predictable performance behavior;
seemingly reasonable access advice can still introduce unexpected costs by
changing internal execution paths.

Random-access advice provides a direct example. An application may choose
file offsets randomly while each call still reads a large contiguous region.
In the fio configuration used in our experiments, random reads issue
\texttt{POSIX\_FADV\_RANDOM} advice by default~\cite{FioFadviseHint}. This
choice appears reasonable, yet in a controlled comparison on the same
instance, simply disabling the advice increased Linux read bandwidth from
734.2~MiB/s to 1413.8~MiB/s, or 1.93$\times$ the original bandwidth. Choosing
parameters from the name of an access pattern is therefore insufficient: the
application requested the same data on the same hardware, but paid a
substantial performance penalty for seemingly reasonable advice.

This observation reveals a more general difficulty: hardware capability does
not automatically translate into performance obtained by the application. A
similar throughput plateau may arise from host-side work, insufficient
request supply, or a device that has already reached its processing plateau,
and each cause demands a fundamentally different intervention. As device
capability grows and internal organization becomes more complex, this problem
will become even more pronounced: a fixed software path may exploit only a
fraction of the available hardware capability, and a hardware upgrade may
fail to improve application performance. Our experiments further show that
fully utilizing an SSD does not necessarily require asynchronous I/O
interfaces with more complex programming models, such as \texttt{io\_uring},
more submission threads, or a more powerful machine. In the scenarios we
measured, a single submission executor already supplied the device fully when
reads were issued through the synchronous \texttt{read} system call. What
matters is understanding how work is organized, how requests reach the
device, and how device completions translate into application progress.

These seemingly scattered optimization experiences and counterintuitive
negative results share a common source and admit a unified explanation.
Application performance is not determined by any isolated layer; it is the
joint result of host-side work, request supply, device-completion behavior,
and the overlap among them. Without understanding this process, final metrics
such as throughput or latency alone make it difficult both to identify the
bottleneck and to choose an optimization reliably.

To establish when these optimization experiences apply, we need a measurable
and testable analytical framework. It must locate what limits read performance
and distinguish whether an intervention reduces work, increases processing
capacity, or overlaps existing work with another stage. We investigate three
questions. Which host-side work remains exposed on the read critical path,
which problems have existing Linux mechanisms addressed, where are their
boundaries, and how much room remains for shortening the critical path? For a
device whose internal mechanisms are hidden, can a small number of
independently calibratable quantities describe the completion process
observed by software and explain how those external quantities arise from
internal mechanisms? How does device-completion behavior propagate through
the software stack into software waiting, and can it then explain, predict,
and guide different software decisions?

For host-side work, we begin with an end-to-end path decomposition and
distinguish work volume, processing time, and critical-path exposure. Linux
uses large folios to reduce the number of metadata objects and readahead to
overlap I/O with copying, but the effects depend on access advice, object
granularity, cache state, and execution organization. We use out-of-order
early copy to expand the available overlap opportunities, parallel copy to
shorten residual work that overlap cannot cover, and the same parallel
framework for submission paths on synchronous backends including BRD, zram,
and emulated PMEM. The results show that parallelism shortens the critical path
only when its executors can distribute actual host-side work; for an SSD that
a single executor already supplies fully, faster submission does not improve
the complete read.

For device-side waiting, we abstract the first-completion wait $F$ and
subsequent completion capability $B_{CQ}$ from software-visible request-batch
completion timelines. We study how they vary with request size, batch size,
and access state, and combine real-device measurements with internal simulator
events to explain their mechanism-level origins. We further distinguish
finite-batch subsequent completion capability from the sustained-supply
plateau and construct a cross-layer wait chain along the actual
folio--bio--request mappings. This chain connects device completions to the
block layer, filesystems, and Linux readahead, while preserving batch overlap
and data-demand order in its predictions.

Finally, we test the practical value of the model and wait chain through four
classes of software decisions on buffered-read and controlled direct-I/O
paths. By distinguishing independent prediction from post-hoc attribution and
comparing model recommendations with actual interventions, we form a closed
loop of observation, modeling, understanding, and control, providing explicit
grounds for applying empirical optimizations. We further recalibrate the
device parameters independently on the higher-specification KI2 instance.
Although the absolute values of $F$ and $B_{CQ}$ change substantially, the
request-batch completion structure, cross-layer propagation, and the main
conclusions of two decision cases are reproduced; Appendix~\ref{app:ki2-replication}
reports the complete results.

The main contributions of this paper are as follows:
\begin{enumerate}
\item \textbf{Read-path analysis and host-side interventions.}
We distinguish work volume, CPU processing time, and critical-path exposure,
and reveal how large folios and I/O--copy overlap depend on access advice,
cache state, and backend paths. Controlled interventions through out-of-order
early copy, opportunistic parallel copy, and parallel submission to
synchronous backends determine how much more aggressive overlap and
parallelism can shorten the critical path and identify their applicability
boundaries.

\item \textbf{A calibratable device-completion model.}
From request-batch completion timelines, we abstract the first-completion wait
$F$ and subsequent completion capability $B_{CQ}$, which respectively capture
when the first completion becomes visible and how the remaining requests in
the batch complete. Real-SSD measurements, internal events from MQSim, and
one-factor interventions reveal their trends and mechanism-level origins.

\item \textbf{A cross-layer wait chain.}
We follow folio--bio--request mappings to track request splitting, merging,
and regrouping, and identify propagation boundaries including interrupts,
tags, the scheduler, and filesystem metadata. Given the request stream that
actually reaches the device, independently calibrated device parameters
predict changes in the waiting observed at the block layer and at
\texttt{read\_pages}.

\item \textbf{Validation through software decisions.}
Four use cases test the framework by deciding when SSD parallel submission is
worthwhile, deriving the waiting range and selection conditions of Linux
readahead windows, predicting the benefits of combining dependent reads and
layout choices under conditional access, and using the first-completion wait
to design polling and sleeping policies. They cover benefits, no benefit, and
benefit reversals.
\end{enumerate}

We plan to release the measurement tools, experimental configurations, data,
and analysis scripts supporting our main findings to enable reproduction and
future research.

\section{Experimental Setup}
\label{sec:methodology}

This paper primarily studies synchronous, large-buffer buffered reads in
Linux. Unless otherwise stated, experiments read local files through
synchronous \texttt{read}- or \texttt{pread}-like interfaces. Experiments that
isolate device behavior use direct I/O, those that study synchronous backends
use BRD or zram, and analyses of internal device mechanisms use
MQSim~\cite{tavakkol2018mqsim}. Path-specific methods and parameters are
introduced when first used.

\textbf{Experimental platforms.} We reproduce the main performance results on
a Huawei Cloud KI1 instance and validate key findings on a KI2 instance. Our
primary platform is a \texttt{ki1.2xlarge.4} instance with eight Kunpeng 920
vCPUs, 32~GiB of memory, and an approximately 3.2-TiB local NVMe SSD;
cross-platform validation uses a 16-vCPU KI2. Both instance types use one local
NVMe SSD, XFS, and Linux 6.6.58-based kernels. Each direct
comparison is performed on the same instance and in the same environment. All
fio experiments use fio 3.39.

\textbf{Common workloads and controls.} Each comparison reads the same ranges
of prewritten files in the same order. Unless otherwise stated, every
cold-cache run executes \texttt{sync} and drops the page cache before
measurement. Tests run serially, without other benchmarks or performance
analysis running concurrently. Interrupt handling, CPU affinity, and polling
are reported when they may affect a conclusion and are held constant within
each direct comparison.

\textbf{Measurement and reporting.} We compute read bandwidth from the amount
of data returned to the application and end-to-end time, treating system-call
completion as the application-visible endpoint of a buffered-read experiment.
Normal performance runs do not load high-frequency probes; internal-path
analyses use separate instrumented runs and compare them with probe-free
results. Observation boundaries are defined when the corresponding data are
first used. Unless otherwise stated, end-to-end and host-side configurations
run independently ten times, while key device-completion-time distributions
use 30 runs; figures report the summary statistics and uncertainty ranges.

\section{Read-Path Decomposition and Metadata Management}
\label{sec:findings}

\begin{figure*}[t]
  \centering
  \includegraphics[width=\textwidth]{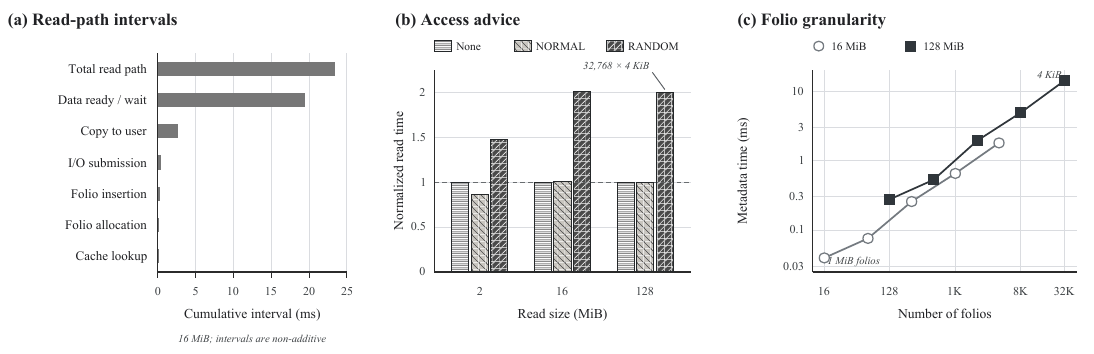}
  \caption{Read-path costs and folio granularity.
  (a) Non-additive cumulative function intervals for a diagnostic 16~MiB cold
  read; data-ready/wait is not pure device time.
  (b) Probe-free read time normalized to no advice at each size.
  (c) Metadata time at fixed data volume across folio sizes.}
  \label{fig:metadata-management}
\end{figure*}

Figure~\ref{fig:metadata-management}(a) groups the main work and waiting in a
16~MiB cold read by role: metadata management, represented by folio allocation
and XArray operations; data movement into the user buffer; and I/O waiting
while required data are not ready. The buffered-read path must handle these
components, but their work volume and execution can change. This section
examines how Linux reduces metadata costs and how this capability depends on
access advice and readahead parameters; subsequent sections address data
movement and device waiting.

Access advice provides a counterintuitive example. We issue reads at random
file offsets, but each call still accesses a contiguous range, so declaring
random access is not semantically unreasonable; in the fio~3.39 version we
use, \texttt{randread} also issues \texttt{POSIX\_FADV\_RANDOM} advice by
default. However, for a 128~MiB cold read, both issuing no advice and
explicitly using \texttt{POSIX\_FADV\_NORMAL} take approximately 186~ms,
whereas \texttt{POSIX\_FADV\_RANDOM} takes approximately 373~ms, nearly twice
as long (Figure~\ref{fig:metadata-management}(b)). Individual 16~MiB reads and
random-offset reads totaling 1~GiB show the same trend.

Further analysis shows that this degradation accompanies a substantial
increase in page-cache objects. For a 128~MiB read, the normal path creates
about a thousand folios, mostly 128~KiB, whereas \texttt{RANDOM} represents
the same range with 32,768 4~KiB folios and increases allocation, insertion,
and per-folio work. The reading thread's CPU time rises from approximately
24.0~ms with no advice to 52.5~ms with \texttt{RANDOM}. To isolate the role of
object count, a controlled in-kernel microbenchmark fixes the logical data
volume and varies folio granularity. Managing 128~MiB with 32,768 4~KiB folios
costs approximately 14.24~ms in metadata operations; using 128 1~MiB folios
reduces this to 0.277~ms, with metadata time clearly correlated with folio
count across configurations (Figure~\ref{fig:metadata-management}(c)). Equal
data volumes therefore need not imply equal management work: page-cache object
granularity directly affects its cost. Metadata work explains part of the
additional host cost, while the complete read-time difference also involves
data movement and I/O waiting. In the measured 16~MiB and 128~MiB NVMe cold
reads, large folios in the normal no-advice path keep cumulative metadata cost
to only about 2.3\%--2.5\% of complete read time.

Whether Linux uses large folios also depends on the readahead window. For a
128~MiB cold \texttt{pread} at a nonzero offset, increasing the window from
2~MiB to 128~MiB raises read time from 50.06 to 88.23~ms and changes the folio
shape from 231 large folios to 32,768 4~KiB folios. Reading-thread CPU time
simultaneously rises from 25.44 to 49.95~ms. Source inspection and path tracing
show that without sequential-access history, a request no larger than the
window may be classified as an isolated random demand and enter a path that
allocates order-0 folios individually. Reading from the beginning of the file,
first reading the immediately preceding page through the same file descriptor,
or making the request larger than the window restores the large-folio path in
the tested configurations. The window's effect therefore also depends on
request size, starting offset, and access history; a larger window does not
necessarily lower metadata cost.

Overall, Linux uses large folios to reduce page-cache object count and keep
metadata costs low, but applications must use access advice and readahead
configuration appropriately to realize this capability. Randomness across
calls does not replace considering contiguous demand within each call, and a
window size cannot be mechanically equated with the application's largest
read.

\section{Data Movement: Overlap and Parallelism}
\label{sec:host-work}

The previous section showed that Linux can substantially reduce metadata costs
by using large folios to manage fewer objects. Data movement is different:
delivering data to a user buffer still requires copying the corresponding
bytes, and its work volume depends primarily on the amount of data read rather
than the number of folios. In our measurements, copy is the largest cost other
than I/O waiting, and increasing folio size cannot reduce it in the same way as
metadata-management work.

Linux reduces the visible cost of copying by overlapping it with device I/O.
When some folios are ready while other submitted data are still being processed
by the device, the reading thread can copy the former, allowing host-side data
movement and device processing to proceed concurrently. This mechanism is
highly effective in the measured reads: during a 128~MiB cold read, target I/O
was in flight throughout the copy calls that moved 98.1\% of the bytes; the
corresponding fraction was 96.1\% at 16~MiB. Linux thus schedules most data
movement while the device is processing I/O. This fraction describes bytes
whose copying overlaps I/O; it does not mean that the same fraction of
end-to-end time is eliminated. The cumulative time of an operation is not its
contribution to end-to-end time.

\subsection{Expanding Overlap with Out-of-Order Early Copy}
\label{sec:out-of-order-copy}

Although this mechanism is effective for cold reads, we find that its effect
depends on cache state and the locations of cache hits. With a fixed 50\% hit
ratio and alternating 2~MiB hit and miss regions, the fraction above falls to
42.8\% for a 128~MiB read. When all cache hits are placed in the latter half of
the requested range, it is 49.0\%. The main reason is that Linux advances copy
in file-offset order: while data near the beginning are not ready, the reading
thread cannot skip over them to copy later data that are already cached.

To exploit these opportunities lost to the ordering constraint, we implement a
more aggressive out-of-order early-copy mechanism in MARS, an experimental
in-kernel prototype used for controlled interventions. While waiting for data
from the device, MARS
checks the state of each folio. If a folio is a ready cache hit, or completes
I/O during the wait and passes the data-validity check, MARS copies it to the
corresponding location in the user buffer according to its file offset, without
waiting for all preceding folios to become ready in order.

To preserve the continuous-prefix semantics of \texttt{read} while copying data
out of order, MARS separates physical data movement from logical read progress.
For early copy, it constructs a temporary \texttt{iov\_iter} from the folio's
file offset to locate the target region in the user buffer, but neither advances
the main \texttt{iov\_iter} nor changes the file position recorded in the
\texttt{iocb}; it also records copy progress within the set of cache-hit folios.
In the final Copy stage, MARS still examines, in file-offset order, the
continuous successful prefix that can be returned. For a folio already copied,
this stage advances only the main iterator, file position, and completed-byte
count, without copying the data again; a ready folio not copied early follows
the normal copy path. Physical copying therefore need not follow file-offset
order, while the system-call return value and file position are still committed
as a continuous file prefix. The complete protocol, including user-buffer
behavior when lower-level I/O returns an error, is provided in
Appendix~\ref{app:early-copy-errors}.

\begin{figure*}[t]
  \centering
  \includegraphics[width=0.72\textwidth]{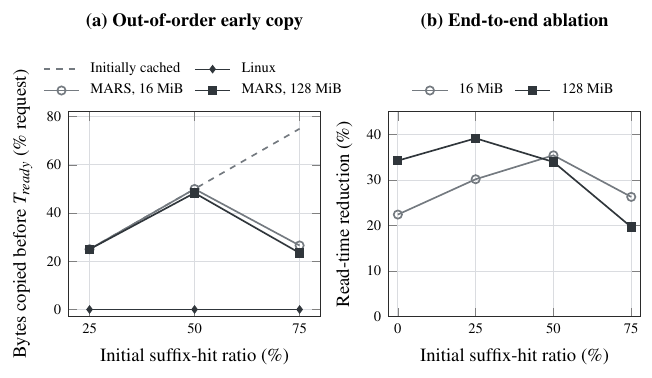}
  \caption{Out-of-order early copy and its end-to-end effect.
  (a) Fraction of the request copied from the initially cached suffix before
  all misses become ready ($T_{\mathrm{ready}}$); copy calls crossing this
  boundary are excluded from the definite fraction.
  (b) Reduction in complete-read time when early copy is enabled within MARS,
  relative to the otherwise identical early-copy-off configuration.}
  \label{fig:out-of-order-copy}
\end{figure*}

We fix the read size at 16 and 128~MiB, vary the initial cache-hit ratio from
0\% to 75\%, and place all hits at the end of the requested range. Linux uses
its native path; within the prototype, all settings other than early copy are
held constant. User buffers are prepared before timing. From independent
completion events and copy timelines, we measure the amount copied before all
target data become ready and separately identify bytes copied from the initial
cache-hit set. Copy intervals crossing this boundary are excluded from the
definite early-copy amount. The suffix-hit experiment directly exposes the
effect of out-of-order execution (Figure~\ref{fig:out-of-order-copy}(a)). For a
128~MiB read with a 50\% suffix hit ratio, Linux copies none of the initially
cached 64~MiB suffix before all misses in the prefix become ready, whereas MARS
copies at least 61.875~MiB of it early. Linux's zero here applies only to the
initially cached suffix; it does not mean that Linux fails to copy completed
I/O from the prefix while waiting for other data. This comparison asks whether
ready suffix data can be used early. The advantage depends on the locations of
cache hits and does not imply that Linux generally lacks overlap. Moreover, as
the hit set grows and fewer misses remain, the device-processing window
available for copying may shrink. The amount of work eligible for early
execution and the overlap time available to execute it are two distinct
constraints.

The byte timeline shows that out-of-order execution can use cached folios that
would otherwise be blocked by unfinished data near the beginning. To verify
that this change in execution also translates into an end-to-end benefit, we
toggle only early copy within MARS. For a 128~MiB read with a 50\% suffix hit
ratio, enabling early copy reduces complete-read time by 34.0\%
(Figure~\ref{fig:out-of-order-copy}(b)). This benefit includes early processing
of both initially cached folios and folios that become ready during the wait;
it therefore cannot be attributed entirely to the cached suffix shown in the
figure.

\subsection{Parallelizing Residual Copy without an Overlap Window}
\label{sec:parallel-copy}

Hiding copy behind device waiting is a reasonable and effective Linux design,
but it has a fundamental prerequisite: the read path must contain a sufficient
device-waiting window. We examine two cases in which this prerequisite no
longer holds. First, a fully warm read issues no data I/O but still has to copy
all requested data to the user buffer. Second, memory backends such as BRD and
non-DAX PMEM complete data processing synchronously in the submission context
and do not leave a separate device-waiting window like an asynchronous SSD.
In both cases, copy can no longer overlap device waiting and its time is exposed
directly on the read path. With no waiting window to exploit, we instead use
more aggressive parallel execution to shorten the residual Copy stage.

The reading thread and opportunistic MARS workers cooperate on this work. MARS
partitions the bytes that remain to be copied into non-overlapping user-buffer
ranges, respecting folio boundaries when assigning tasks. The reading thread,
acting as the master, executes one share, while currently available workers
receive the others. Data copied early are excluded from the remaining movement.
A worker temporarily uses the application's address space, and the user pages
are made writable before execution. If no worker is available, a task is not
accepted, or a worker yields its CPU partway through, the master resumes from
the recorded progress and completes the remainder. Ownership and progress
records prevent the master and workers from processing the same task
concurrently, and the read returns only after all work completes. Workers thus
provide opportunistic acceleration, while the master retains responsibility
for completing the read. The complete master--worker protocol, ownership
transitions, and limitations of this execution framework are given in
Appendix~\ref{app:worker-protocol}.

In addition to a fixed total executor count $k$, we evaluate an Auto policy.
Auto first determines an operator-specific upper bound from the remaining work
and a minimum-benefit threshold for adding an executor, and then selects the
actual degree of parallelism from the workers available at decision time and
the number of online CPUs. It therefore treats currently available CPUs as a
shared resource and bounds the executor count without solving for the
runtime-optimal degree of parallelism for each operation. Scans over fixed $k$
provide performance references, while Auto shows how the same framework can
select parallelism according to available resources. The selection formula and
its parameters are provided in Appendix~\ref{app:auto-selection}.

\begin{figure*}[t]
  \centering
  \includegraphics[width=\textwidth]{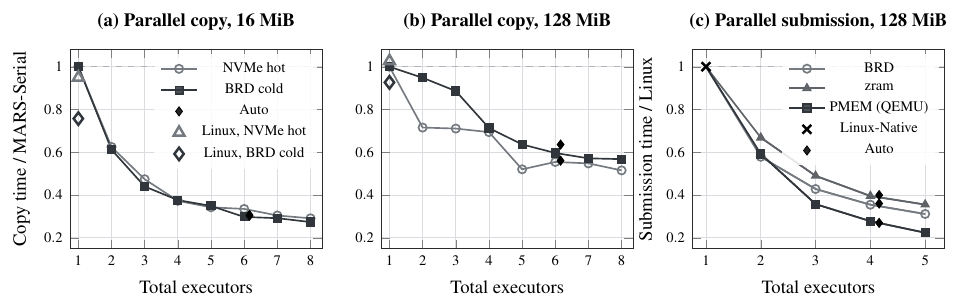}
  \caption{Parallel host-side work. In (a) and (b), 1 denotes MARS-Serial
  for the corresponding fully warm NVMe-page-cache or BRD-cold condition;
  the hollow triangle and diamond show the explicitly labeled Linux-Native
  references, not the speedup denominators. In (c), 1 denotes native Linux
  separately for each backend. The three
  native baselines coincide at the single cross at $(1,1)$, and each line connects
  this point to MARS Parallel-2 through Parallel-5. Filled diamonds show Auto
  at its median selected count.
  PMEM is emulated with QEMU and is compared only within that backend. All
  values are operator times rather than complete buffered-read times.}
  \label{fig:parallel-host-work}
\end{figure*}

For fully warm and BRD cold reads, we compare Linux-Native, MARS-Serial, fixed
$k=2\ldots8$, and MARS-Auto. All MARS configurations enable early copy and
prepare the user buffer before the read; the principal intervention across
MARS configurations is therefore the number of executors in the final Copy
stage. Each configuration has ten uninstrumented reads and ten independent
stage-diagnostic runs, and every main experiment passes full-data validation.
MARS stage time includes task dispatch, worker and master execution, fallback,
and joining. The cumulative time of native Linux copy functions is shown as a
reference and is not used as the denominator of the parallel speedup.

Figure~\ref{fig:parallel-host-work}(a--b) presents the parallel-copy results.
For a
fully warm 16~MiB read, Parallel-4 shortens final copy from 3.378 to 1.260~ms,
a 62.7\% reduction; Auto shortens it to 1.035~ms, a 69.4\% reduction. BRD cold
reads show a similar effect: at 16~MiB, Parallel-4 and Auto reduce this stage by
62.4\% and 69.6\%, respectively. Both conditions still benefit at 128~MiB, but
scaling is not linear: Parallel-4 reduces stage time by 30.6\% for fully warm
reads and 28.6\% for BRD, while Auto reduces it by 43.9\% and 36.4\%. These
results also show that a higher degree of parallelism does not necessarily
produce a larger benefit.

These results show that overlap and parallelism address different conditions.
While the device is still processing data, overlap can reduce the portion of
copy that separately occupies the read timeline. When that opportunity is
absent, parallelism can shorten the residual Copy stage. The two mechanisms do
not substitute for each other, and neither implies fewer bytes moved or less
total CPU work.

\subsection{Applying the Same Parallel Method to Synchronous Submission}
\label{sec:parallel-submission}

Following the same reasoning, we next examine I/O submission. Parallel I/O
submission reuses the master--worker framework and resource-selection policy
above: the master still participates in execution and retains final completion
responsibility, and workers remain opportunistic collaborators. The unit of
work changes from a range of data to copy to a complete readahead batch, and
the operator-specific bound is instead determined by the number of batches
that can be processed independently. MARS partitions folios awaiting
submission at complete-readahead-batch boundaries and assigns different
batches to the master and workers, allowing several executors to construct and
submit read requests concurrently. On a backend that completes requests
synchronously in the submission context, this distributes inline backend work,
such as decompression and memory copying, across executors.

Because an opportunistic worker cannot enter a potentially sleeping path while
executing an operator, MARS augments Linux's existing \texttt{IOMAP\_NOWAIT}
and \texttt{REQ\_NOWAIT} semantics with a handshake between a request-side flag
and a queue capability flag. A worker attempts submission only when the backend
explicitly supports the corresponding nonblocking path. Each worker maintains
an independent cursor and progress context, recording its folio range, actual
submission progress, the first unfinished position, and the folios it still
holds; together these form the recovery context for the submission operator.
If an operation cannot complete immediately, the worker returns the remaining
range to the master with the first unfinished folio as the takeover boundary.
The master continues from that boundary through the normal sleepable path,
without resubmitting I/O that has already been issued successfully.

Synchronous backends such as BRD, zram, and non-DAX PMEM do not hand requests
to an independent device for asynchronous progress. Data movement or
decompression that would otherwise form part of device service is instead
performed synchronously by the submitting CPU. Submission therefore does more
than supply requests to a backend; it carries out actual processing on the
read path. In a path breakdown of native Linux 128~MiB cold reads, the
cumulative time in the BRD and zram driver submission functions accounts for
approximately 52.9\% and 58.3\%, respectively, of diagnostic end-to-end read
time. Because these backends complete the actual processing synchronously in
the submission context, this time is exposed directly on the critical path and
forms the dominant part of end-to-end time.

To isolate this effect, we begin timing after inputs such as folios have been
prepared and compare only the submission stage. At 128~MiB on native KI1,
serial Linux submission takes 27.428~ms for BRD, compared with 8.545~ms using
five executors and 9.885~ms with Auto; for zram, the corresponding times are
29.818, 10.600, and 11.925~ms. Auto usually selects four total executors in
these two cases and shortens the submission stage by approximately 64.0\% and
60.0\%, respectively. To test whether the same method applies to another
synchronous memory-backend path, we use an emulated non-DAX PMEM device in a
QEMU environment on KI1. On this path, Auto likewise shortens the 128~MiB
submission stage by 73.0\% (Figure~\ref{fig:parallel-host-work}(c)).

Overall, Linux already hides most copy cost in common cases by overlapping it
with device waiting, but this approach is constrained by the available waiting
window and the order in which data become ready. Out-of-order early copy
expands the overlap opportunities that can be used; when no waiting window
exists, parallel copy shortens the data-movement stage exposed on the read
path. The same opportunistic parallel framework also distributes actual work
performed in the submission context of synchronous backends, covering
host-side work that overlap cannot address.

\section{Device Completion Model: From Completion Timelines to Internal Mechanisms}
\label{sec:model}

The previous section showed that additional executors can distribute the host processing performed in the submission context of synchronous backends such as BRD, zram, and emulated PMEM. On an SSD, however, the same approach shortened request submission without improving complete-read performance. To understand why, we next examine the per-request completion behavior exposed by the SSD at the driver layer.

\subsection{Separating Two Components from Completion Timelines}

To isolate upper-layer effects such as the file system and page cache, we submit finite batches of equal-sized requests directly to the NVMe device and record the submission and completion time of every request. In each run, a thread pinned to one CPU submits requests consecutively, while another CPU independently polls the completion queue. Unless otherwise stated, each run scans fresh, nonoverlapping address ranges.

\begin{figure*}[t]
  \centering
  \includegraphics[width=\textwidth]{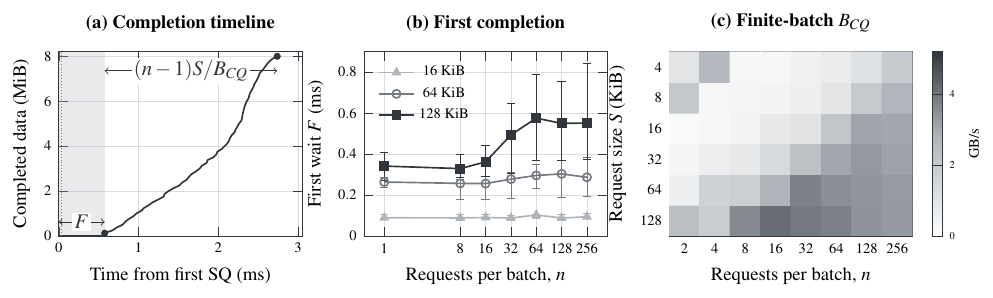}
  \caption{Host-visible completion behavior on the KI1 SSD. (a) Median cumulative completion timeline over 30 runs for a $128\,\mathrm{KiB}\times64$ batch; the shaded interval is $F$. (b) $F$ as request size and batch size vary; error bars show P10--P90. (c) The corresponding $B_{CQ}$ matrix.}
  \label{fig:device-completion-model}
\end{figure*}

Figure~\ref{fig:device-completion-model}(a) shows the cumulative completion curve of a $128\,\mathrm{KiB}\times64$ batch on the KI1 SSD. The curve has a clear two-stage structure. After the first request is submitted, the host initially waits without observing a completion; after the first completion arrives, the remaining requests complete densely over a subsequent interval. The other request sizes we tested exhibit the same completion-timeline structure. We therefore decompose the completion time of a finite batch into a \emph{first-completion wait} and a \emph{subsequent completion process}.

Consider a batch of $n$ equal-sized requests, each of size $S$. Let $t_0$ denote the time at which the first request enters the submission queue, and let $c_1$ and $c_n$ denote the times at which the host observes the first and last completions. We define the first-completion wait as

\begin{equation}
F=c_1-t_0,
\label{eq:first-completion-wait}
\end{equation}

and the effective finite-batch completion capability after the first completion as

\begin{equation}
B_{CQ}=\frac{(n-1)S}{c_n-c_1},\qquad n\geq2.
\label{eq:bcq}
\end{equation}

The total completion time of the finite batch is thus

\begin{equation}
T=c_n-t_0=F+\frac{(n-1)S}{B_{CQ}}.
\label{eq:finite-batch-model}
\end{equation}

Here, $F$ is the host-visible wait from the first submission until the first observed completion; it is not the NAND read latency. $B_{CQ}$ describes only the effective completion capability between the first and last completion of a finite batch, and does not assume that intermediate requests complete at uniform intervals. When $F$, $T$, and $B_{CQ}$ come from the same batch, Equation~\ref{eq:finite-batch-model} is an identity. Its value depends on whether the two quantities exhibit calibratable regularities, whether device mechanisms can explain them, and whether independently obtained parameters can predict new batches.

\subsection{External Behavior of \texorpdfstring{$F$ and $B_{CQ}$}{F and BCQ}}

We next vary request size $S$ and batch size $n$. Figure~\ref{fig:device-completion-model}(b) shows that $F$ generally increases with request size, whereas batch size has a milder effect at a fixed request size. Across the measured batch sizes, $F$ is approximately 0.09--0.10\,ms for 16\,KiB requests and increases from approximately 0.341 to 0.552\,ms for 128\,KiB requests. The P10--P90 ranges further show that $F$ is a distribution under a fixed protocol and device state, rather than an identical constant in every run.

The subsequent completion capability follows a different pattern. Figure~\ref{fig:device-completion-model}(c) presents the $S\times n$ matrix on KI1. Despite local peaks and nonmonotonicity, $B_{CQ}$ generally increases toward larger requests and larger batches, approaching a common completion capability in the large-request, large-batch region. For short batches, the endpoint interval is affected by work completed before the first host completion, completion clustering, and final draining. A locally high $B_{CQ}$ therefore does not directly imply a higher sustained device bandwidth.

These measurements show that both $F$ and $B_{CQ}$ depend on request shape. We therefore write the model more explicitly as

\begin{equation}
T(n,S)=F(n,S)+\frac{(n-1)S}{B_{CQ}(n,S)}.
\label{eq:shape-dependent-model}
\end{equation}

To determine the capability toward which large batches converge, we run a replenishment experiment that maintains a target number of outstanding requests and computes the stable rate $B_{\infty}$ over the middle 80\% of completions. On KI1, requests of 8\,KiB and larger reach a plateau of approximately 3.44--3.57\,GB/s at sufficient queue depth; both 64\,KiB and 128\,KiB requests reach approximately 3.57\,GB/s. Requests of 512\,B and 4\,KiB do not satisfy our pre-established plateau criterion within the tested queue-depth range.

$B_{CQ}(n,S)$ describes a finite batch that starts and eventually drains, whereas $B_{\infty}$ describes the processing plateau under sustained supply after excluding startup and draining. The two must be measured separately. An independently measured $B_{\infty}$ can approximate $B_{CQ}(n,S)$ only when request granularity and batch size have already brought the subsequent completion capability close to the plateau.

\subsection{From External Parameters to Internal Mechanisms}

\begin{figure*}[t]
  \centering
  \includegraphics[width=0.94\textwidth]{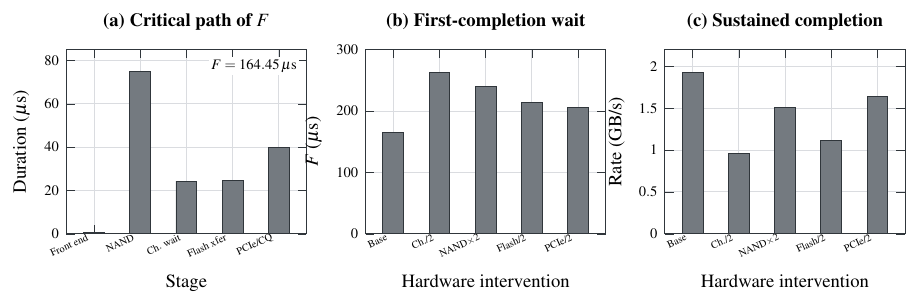}
  \caption{MQSim evidence for the internal mechanisms behind the external parameters. (a) Critical-path decomposition of $F$ for a single 128\,KiB request. (b) $F$ under one-factor hardware interventions. (c) Sustained host-visible completion capability under the same interventions.}
  \label{fig:mqsim-internal-mechanism}
\end{figure*}

A real SSD exposes only submission and completion endpoints to the host. These endpoints alone cannot reveal which internal processes create $F$ and the subsequent completion capability. We use MQSim to answer two questions: whether the same two-stage completion structure emerges in a device model whose internal events are fully visible, and which internal stages and resources jointly shape the two external parameters. We first hide the simulator's internal events and retain only the same host submission and completion endpoints available on the real device. We then enable internal tracing and apply one-factor interventions.

Across the tested $S$ and $n$ configurations, MQSim exhibits a completion-timeline structure similar to the KI1 SSD. For the same $128\,\mathrm{KiB}\times64$ configuration used in Figure~\ref{fig:device-completion-model}(a), $F$, $T$, and $B_{CQ}$ are approximately 164.95\,$\mu$s, 4.49\,ms, and 1.91\,GB/s, respectively. Thus, different internal device organizations can produce the same two-stage external structure.

Internal events show that $F$ is not a single hardware latency, but the critical path traversed by the internal transaction that determines the first host completion. In the single-request baseline, one 128\,KiB request is split into sixteen 8\,KiB internal transactions and distributed across eight channels. Their NAND reads overlap extensively, but data outputs on the same channel remain serialized. For the internal transaction that determines the first host completion, the critical path is

\begin{equation}
\begin{aligned}
F={}&t_{\mathrm{H2D}}+t_{\mathrm{sched}}+t_{\mathrm{cmd}}+L_{\mathrm{NAND}}\\
   &+t_{\mathrm{channel\ wait}}+t_{\mathrm{data}}+t_{\mathrm{return}}.
\end{aligned}
\label{eq:f-internal-path}
\end{equation}

The terms denote host-to-device transfer, internal queuing and scheduling, Flash command transfer, NAND read latency, waiting for data output on the same channel, Flash data transfer, and PCIe and completion return after the data become ready. Figure~\ref{fig:mqsim-internal-mechanism}(a) shows the measured decomposition. In addition to approximately 75\,$\mu$s of NAND read latency, the path includes channel-output waiting, Flash data transfer, and PCIe and completion return, yielding an $F$ of approximately 165\,$\mu$s. Thus, $F$ is a host-visible quantity jointly produced by multiple internal stages and their overlap, rather than an isolated hardware latency. This decomposition describes the critical path in our MQSim configuration and does not assume that the real KI1 SSD has the same internal organization.

Internal events also reveal the meaning of $B_{CQ}$. Let the total batch size be $N=nS$, and let $X_F$ be the total number of bytes whose Flash data transfer has completed by the time the first host completion arrives. Excluding the $S$ bytes of the first completed request, the work already completed for other requests before $F$ is

\begin{equation}
W_F=X_F-S.
\label{eq:prework}
\end{equation}

The finite-batch endpoint completion capability can then be decomposed as

\begin{equation}
\begin{aligned}
B_{CQ}
&=\frac{(n-1)S}{T-F}\\
&=\frac{N-X_F}{T-F}+\frac{X_F-S}{T-F}\\
&=\frac{N-X_F}{T-F}+\frac{W_F}{T-F}.
\end{aligned}
\label{eq:bcq-internal-accounting}
\end{equation}

Of the 1,648 finite-batch simulations we ran, 1,612 define $B_{CQ}$. The event-accounting relation above holds for every such configuration, with a maximum residual of only $4.44\times10^{-16}$\,GB/s. We observe $W_F>0$ in 1,356 runs (84.1\%), showing that work completed before the first host completion is not a rare boundary case. Subtracting $F$ from total time does not remove the internal work performed for other requests before the first completion; that work subsequently appears as denser host completions. Consequently, $B_{CQ}$ is an endpoint-equivalent quantity formed by internal service, the first-completion boundary, completion clustering, and final draining. It is not the physical bandwidth of a particular hardware link and, for a short batch, can even exceed the sustained plateau.

One-factor interventions further validate these relationships. Figures~\ref{fig:mqsim-internal-mechanism}(b) and~\ref{fig:mqsim-internal-mechanism}(c) show that reducing the number of Flash channels from eight to four increases $F$ to approximately 263\,$\mu$s and reduces sustained completion capability from approximately 1.924 to 0.963\,GB/s. Increasing NAND read latency and reducing Flash or PCIe transfer capability alter the two external quantities in different ways. A separate fetch-window experiment further shows that external request supply does not automatically become internal parallelism. Even while the host maintains 64 outstanding requests, increasing Queue Fetch Size from 1 to 16 raises sustained bandwidth from 0.547 to 1.925\,GB/s, after which a larger window provides no further gain. Requests in the host submission queue contribute to device completion capability only after the controller fetches and organizes them into parallel internal work.

Taken together, these results yield a device-completion model comprising $F$ and $B_{CQ}$, abstracted from the completion-timeline structure of request batches. Controlled experiments on a real SSD reveal distinct and calibratable trends in both parameters as request size $S$ and batch size $n$ vary. We further reproduce the same completion structure in MQSim and use internal event tracing and one-factor interventions to explain the device mechanisms that produce $F$ and $B_{CQ}$ and the principal factors that affect them. Thus, the model not only describes the host-observed completion form mathematically, but also connects these external parameters to internal device mechanisms.

\section{The Cross-Layer Wait Chain: From Device Completions to Software Waiting}
\label{sec:wait-chain}

The preceding section considered batches of equal-sized requests that had
already reached the device, and characterized the completion behavior visible
to the host. Applications, however, do not normally construct and issue NVMe
requests directly. The I/O stack spans the page cache, file system, block
layer, and device layer. Each layer represents the same read demand using a
different object---including folios, bios, requests, and NVMe commands---and
applies its own transformations to the request shape and execution path. An
NVMe request is only the device input at the end of this path. After the
device completes it, the completion must travel in the opposite direction
through request and bio completion and folio unlocking before the reading
thread can make progress. In other words, requests are transformed as they
move down the I/O stack, whereas device completions are progressively
translated into software progress as they move upward. We call this path from
device completion to software waiting the \emph{cross-layer wait chain}.

To propagate the device model up this chain, we must answer three questions:
what request stream upper-layer objects ultimately produce, how mechanisms at
each layer alter requests and completion propagation, and whether device
completion behavior can explain and predict the waiting observed above those
layers once these effects are retained. We therefore recover the
folio--bio--request mappings, examine the principal factors at each layer,
and test whether independently measured device-batch times predict waiting at
the block layer and at \texttt{read\_pages}.

\subsection{Recovering Device Requests from Upper-Layer Objects}

The device model consumes the actual request stream, rather than the nominal
number of upper-layer objects. The block layer accepts bios and converts them
into requests for queuing and dispatch. A bio may remain one request or be
split, while compatible adjacent bios may be merged into one request. On the
measured NVMe path, 16, 64, and 128~KiB bios remain one request of the same
size. Bios of 256 and 512~KiB and 1~MiB are instead split into two, four, and
eight 128~KiB requests. Conversely, within one plug, groups of 4, 8, 16, and
32 adjacent 4~KiB bios are progressively merged into one 16, 32, 64, and
128~KiB request. When merging is disabled, they remain independent 4~KiB
requests.

Before this point, the file system accepts folios from the page cache and uses
file offsets and extent mappings to organize contiguous data into bios.
Table~\ref{tab:cross-layer-mapping} gives representative mappings on a
contiguous XFS extent. A single 4--128~KiB folio forms one bio and request of
the same size. For a fixed total of 128~KiB, all six organizations ranging
from 32 4~KiB folios to one 128~KiB folio form the same 128~KiB bio and
request. Larger contiguous inputs are first aggregated by the file system and
then split at the block layer: 64 4~KiB folios form one 256~KiB bio and two
128~KiB requests, whereas eight 128~KiB folios form one 1~MiB bio and eight
128~KiB requests.

\begin{table*}[t]
  \centering
  \small
  \renewcommand{\arraystretch}{1.08}
  \caption{Representative transformations from upper-layer objects to the
  requests issued to the NVMe device.}
  \label{tab:cross-layer-mapping}
  \begin{tabular}{@{}p{0.29\textwidth}p{0.21\textwidth}p{0.22\textwidth}p{0.17\textwidth}@{}}
    \toprule
    Upper-layer input & Bio representation & NVMe-bound requests & Transformation \\
    \midrule
    One 4/16/64/128~KiB folio & One bio of the same size & One request of the same size & One-to-one \\
    32 adjacent 4~KiB bios & 32 4~KiB bios & One 128~KiB request & Block-layer merge \\
    One 256~KiB bio & One 256~KiB bio & Two 128~KiB requests & Block-layer split \\
    32$\times$4 through 1$\times$128~KiB folios; 128~KiB total & One 128~KiB bio & One 128~KiB request & File-system aggregation \\
    64$\times$4~KiB folios & One 256~KiB bio & Two 128~KiB requests & Aggregate, then split \\
    8$\times$128~KiB folios & One 1~MiB bio & Eight 128~KiB requests & Aggregate, then split \\
    \bottomrule
  \end{tabular}
\end{table*}

Both layers organize requests opportunistically. The file system extends the
current bio along a contiguous extent, while the block layer merges adjacent
bios into the current request when their addresses and operations are
compatible and merging is allowed. It also splits bios larger than the queue's
maximum request size, which is 128~KiB in our KI1 configuration. Extent
boundaries, noncontiguous addresses, plug flushes, disabled merging, and size
or segment limits all terminate aggregation; the block layer does not delay
arbitrary requests merely to fill this limit.

These mappings determine the $(n,S)$ supplied to the device model. For
example, the wait for one 1~MiB bio should be compared with
$T(8,128\,\mathrm{KiB})$, rather than with one large request or
$F(1,128\,\mathrm{KiB})$. The measured completion time of the 1~MiB bio is
677.8\,$\mu$s, while the independent eight-request device batch completes in
662.9\,$\mu$s. Cross-layer prediction therefore begins by recovering the
request stream that actually reaches the device, rather than by adding a fixed
software overhead.

\subsection{How Software Changes Wait Propagation}

Recovering the request mapping is not sufficient to determine upper-layer
waiting. Software can also alter request supply, completion processing, and
dependencies that precede the data requests. We use four boundary cases to
summarize these effects.

At the device/driver completion boundary, completion processing can compete
with request supply. For 512 4~KiB requests, median total time is 1062.8,
1150.3, and 1524.4\,$\mu$s with direct device polling, bio polling on a
separate CPU, and ordinary interrupt-driven bio completion, respectively.
The diagnostic run shows that, before I/O submission finishes, the same CPU
spends a cumulative 688.2\,$\mu$s processing IRQs. Completion processing can
therefore slow the supply of subsequent requests in a small-request,
high-count path.

At the block layer, tag shortage changes where waiting occurs. Increasing the
number of 128~KiB requests from 512 to 1536--4096 raises submission time from
0.915~ms to 21.759--115.813~ms, while end-to-end throughput remains
3.34--3.54~GB/s. Waiting for tags substantially stretches submission, but
does not proportionally reduce throughput while the device remains supplied.
The scheduler can also change who dispatches requests. For 256 128~KiB
requests, \texttt{mq-deadline} moves issue work from the submitting thread to
a kworker, while median total time increases from 10.368 to 10.869~ms. This
changes the supply context without creating a different device-completion
behavior.

At the file-system layer, metadata can introduce a new serialized dependency
before the data batch. In the measured cold-extent state, the XFS path issues
three serialized 4~KiB metadata reads before the data requests, adding
282--310\,$\mu$s relative to the metadata-hot state. This time is an extra
prefix before the data batch and cannot be folded into the $F$ or $B_{CQ}$ of
the data requests.

Software therefore does not simply add one fixed constant to device time. By
changing request shape, supply, and completion context, or by introducing new
serialized requests, it determines how device completion behavior becomes
visible as upper-layer waiting.

\subsection{Can Device Batch Time Predict Upper-Layer Waiting?}

After identifying object mappings and within-layer effects, we test whether
the completion behavior and time described by the device model can predict
the waiting observed at software interfaces along the wait chain. We use
\texttt{submit\_bio} at the block layer and \texttt{read\_pages} at the
file-system/page-cache boundary as two observation points, and examine both a
normal single batch and artificially controlled, strictly separated batches.

We first consider a normal \texttt{read\_pages} batch produced by the file
system. \texttt{read\_pages} accepts a batch of folios prepared by the page
cache, maps the file offsets, constructs and submits bios, and unlocks the
folios after the lower-level I/O completes. Its completion time therefore
crosses the file system, block layer, and device layer, making it a natural
point at which to test whether device behavior propagates to folio completion.
With a contiguous XFS extent, hot metadata, and cold data pages, one call
receives $n=8,16,32,64$ contiguous 128~KiB folios. The file system forms one
$n\times128$~KiB bio, which the block layer splits into $n$ 128~KiB requests.
Before observing any \texttt{read\_pages} result, we predict its completion
time directly from the independently sealed device measurement:

\begin{equation}
\widehat T_{\mathrm{read\_pages}}(n)
=T_{\mathrm{dev}}(n,128\,\mathrm{KiB}).
\label{eq:read-pages-prediction}
\end{equation}

\begin{figure*}[t]
  \centering
  \includegraphics[width=0.98\textwidth]{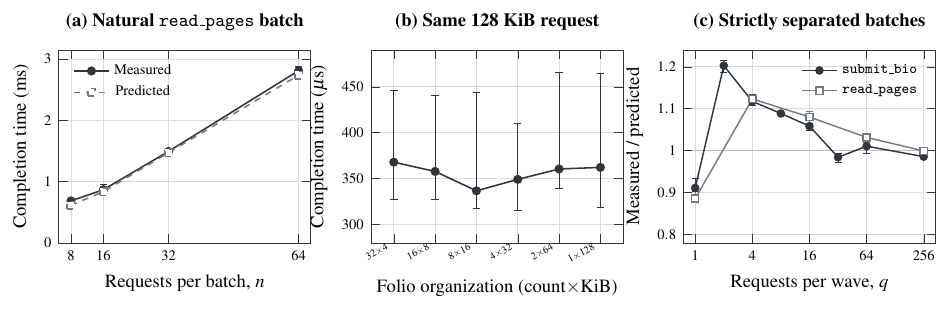}
  \caption{Cross-layer predictions from independently measured device-batch
  times. (a) Predicted and measured \texttt{read\_pages} completion time for
  a normal batch; error bars show P10--P90. (b) \texttt{read\_pages} time for
  six folio organizations mapping to the same 128~KiB request. (c)
  Measured-to-predicted ratio for strictly separated batches at
  \texttt{submit\_bio} and \texttt{read\_pages}. Timing and request mappings
  come from separate uninstrumented and diagnostic runs, respectively.}
  \label{fig:cross-layer-wait-chain}
\end{figure*}

Figure~\ref{fig:cross-layer-wait-chain}(a) compares these predictions with
probe-free measurements. The error is 10.58\% at $n=8$, and 3.69\%, 1.46\%,
and 2.58\% at $n=16,32,64$, respectively. When mapping and preceding
dependencies are controlled, device-batch time thus preserves the principal
magnitude of a normal \texttt{read\_pages} wait.

Identical device input does not imply identical upper-layer time. We fix the
final request stream to one 128~KiB request and vary only its folio
organization, from 32 4~KiB folios to one 128~KiB folio.
Figure~\ref{fig:cross-layer-wait-chain}(b) shows that the median
\texttt{read\_pages} time across the six organizations still spans
336.8--367.9\,$\mu$s. The request stream fixes the common device work, but
upper-layer object organization retains a bounded software difference. The
device model predicts the principal wait, rather than the complete execution
time of every layer.

Finally, we control batch boundaries using strictly separated batches. The
experiment fixes $N=256$ 128~KiB requests, submits only $q$ requests at a
time, and does not submit the next batch until all requests in the current one
complete. Independent device measurements give the prediction

\begin{equation}
\widehat T_{\mathrm{strict}}(q)
=\frac{N}{q}T_{\mathrm{dev}}(q,128\,\mathrm{KiB}).
\label{eq:strict-batch-prediction}
\end{equation}

We instantiate the same batching at two layers. The block-layer microbenchmark
directly submits $q$ one-to-one 128~KiB bios. Each \texttt{read\_pages} call
instead receives $q$ 128~KiB folios, forms one $q\times128$~KiB bio, and lets
the block layer split it into $q$ requests.
Figure~\ref{fig:cross-layer-wait-chain}(c) reports measured time divided by
the independent prediction. Across the eight block-layer configurations, the
median, P90, and maximum absolute errors are 6.83\%, 12.40\%, and 16.84\%,
and the prediction correctly orders all 28 pairs of configurations. Across
the five \texttt{read\_pages} configurations, median and maximum absolute
errors are 7.43\% and 12.88\%. The inter-batch barrier removes overlap
between adjacent batches, so total waiting accumulates multiple device-batch
times. Similar errors at the
\texttt{submit\_bio} and \texttt{read\_pages} endpoints show that the waiting
change induced by batch boundaries crosses the block layer and continues to
folio completion.

Taken together, merging, splitting, and aggregation as requests move downward
determine the actual input to the device model. Mechanisms at each software
layer further determine how device behavior appears on the software critical
path by changing how requests reach the device, altering the completion path,
or inserting additional I/O before the data requests. When these effects are
retained explicitly, independent device-batch times predict the principal
waiting observed at the block layer and at \texttt{read\_pages}, while still
allowing bounded software differences at each layer.

\section{Model-Guided Decisions}
\label{sec:decisions}
To test the practical value of the device-completion model and the wait chain
in more complete software paths, we select four use cases and examine whether
they can explain observed phenomena, predict strategy outcomes, and guide
application decisions.

To make the model independently calibratable on a target device and provide
inputs to subsequent software decisions, we implement a device microbenchmark.
It constructs finite batches of equal-size requests on an NVMe polling queue,
pins submission and completion polling to different CPUs, and records when
each request is written to the SQ and when its CQE is observed by the host.
Each batch drains completely before the next one begins. The first SQ, first
CQ, and last CQ directly yield $F(n,S)$, $T(n,S)$, and $B_{CQ}(n,S)$ for
different request sizes $S$ and batch sizes $n$. Neither the target software
path nor application results participate in parameter fitting. Unless stated
otherwise, the following use cases use device parameters produced by this
microbenchmark.

\subsection{When Is Parallel Submission Worthwhile?}
\label{sec:uc-parallel-submission}

The preceding sections left a counterintuitive result: adding submission
executors on an SSD substantially shortened request submission, but did not
improve the complete read. For a 64~MiB cold read, the two-executor
configuration (P2) reduced concentrated-submission time from 0.867 to
0.500~ms, a $1.73\times$ speedup, while probe-free complete-read time instead
increased from 30.09 to 33.68~ms. Independent path diagnostics confirmed that
P2 used two submission executors and that every target request reached the
driver dispatch path before the complete submission function returned.
Parallel execution therefore accelerated request supply, but this local
speedup did not translate into application performance. Submission time alone
cannot explain why.

We next analyze why parallel submission did not help and under what conditions
it could improve SSD read performance. On the experimental-prototype path,
the complete read demand is known before submission begins, and the reading
thread and workers cooperate to supply the corresponding requests. The CPUs
produce and supply requests while the SSD processes and completes them,
forming an overlapping producer--consumer process. Whether another submission
executor is useful depends on whether software supply is already sufficient
for the device's completion behavior.

The completion model lets us derive this software-supply requirement from the
time needed by the device. Consider a demand that maps to $n$ device requests
of size $S$. Let $B_{\mathrm{submit}}(k)$ be the effective supply capability
of $k$ executors, computed as the total bytes represented by the actual
requests divided by the complete submission-stage time. Supplying the batch
takes $t_{\mathrm{submit}}(k)=nS/B_{\mathrm{submit}}(k)$, whereas the
independently calibrated device model gives

\begin{equation}
T(n,S)=F(n,S)+\frac{(n-1)S}{B_{CQ}(n,S)}.
\label{eq:uc1-device-batch-time}
\end{equation}

The reference batch has an average completion rate of $nS/T(n,S)$. For
submission not to become a bottleneck beyond this reference time, software
must at least supply the entire batch within that interval, i.e.,
$t_{\mathrm{submit}}(k)\leq T(n,S)$. This gives the average supply
requirement

\begin{equation}
\begin{aligned}
B_{\mathrm{submit}}(k)&\geq B_{\mathrm{req}}(n,S),\\
B_{\mathrm{req}}(n,S)
&=\frac{nS}{F(n,S)+(n-1)S/B_{CQ}(n,S)}.
\end{aligned}
\label{eq:uc1-required-supply}
\end{equation}

Among a measured executor set $\mathcal K$, the smallest configuration that
meets this requirement is

\begin{equation}
k_{\min}^{\mathrm{meas}}(n,S)
=\min_{\substack{k\in\mathcal K:\\
B_{\mathrm{submit}}(k)\geq B_{\mathrm{req}}(n,S)}} k.
\label{eq:uc1-kmin}
\end{equation}

$B_{\mathrm{req}}$ is both the reference batch's average completion rate and
the minimum average software-supply threshold. It retains the joint effects
of request size, batch size, first-completion wait, and subsequent completion
capability. This endpoint condition is a necessary screening test: satisfying
it does not ensure sufficient outstanding work at every instant, so the
actual supply timeline and other path bottlenecks must still be checked.

For example, in the independently calibrated $256\times128$~KiB batch,
$F\approx0.552$~ms and $B_{CQ}\approx3.337$~GB/s.
Equation~\ref{eq:uc1-device-batch-time} gives $T\approx10.57$~ms. The batch
contains 32~MiB, so both its average completion rate and the corresponding
average software-supply threshold are approximately 3.18~GB/s. Dividing the
numerator and denominator of Equation~\ref{eq:uc1-required-supply} by $nS$
gives

\begin{equation}
B_{\mathrm{req}}(n,S)=
\left[
\frac{F(n,S)}{nS}
+\frac{1-1/n}{B_{CQ}(n,S)}
\right]^{-1}.
\label{eq:uc1-required-supply-limit}
\end{equation}

As the batch grows, if $F(n,S)/(nS)$ vanishes and $B_{CQ}(n,S)$ approaches
the sustained platform $B_\infty$, then

\begin{equation}
\begin{aligned}
B_{\mathrm{req}}(n,S)&\longrightarrow B_\infty,\\
k_{\min}^{\mathrm{meas}}
&\approx\min_{\substack{k\in\mathcal K:\\
B_{\mathrm{submit}}(k)\geq B_\infty}} k.
\end{aligned}
\label{eq:uc1-sustained-limit}
\end{equation}

For the 64~MiB concentrated read, Serial and P2 supply 77.4 and
134.3~GB/s, respectively, whereas the independently measured sustained
completion platform is only 3.57~GB/s. Both the finite-batch rate of
3.18~GB/s and the large-batch platform of 3.57~GB/s are far below Serial's
supply capability. Even against the higher platform reference, Serial has a
$21.7\times$ supply margin, yielding $k_{\min}^{\mathrm{meas}}=1$.

The P2 result also identifies the boundary at which parallel submission could
matter. Holding the measured software capabilities fixed, Serial becomes the
limiting side only when the device requires more than 77.4~GB/s. When

\begin{equation}
77.4 < B_{\mathrm{req}}(n,S)\leq134.3\ \mathrm{GB/s},
\label{eq:uc1-p2-region}
\end{equation}

P2 can raise insufficient average supply above the model requirement. Above
134.3~GB/s, P2 may still improve supply but remains insufficient itself. The
measured 3.57~GB/s platform is therefore far outside the region that needs
P2. A device with greater aggregate completion capability, or reduced
effective supply caused by CPU quotas or virtualized scheduling, can reverse
this relation and make parallel submission worth reconsidering. Changes in
the device, request shape, submission path, interface layer, CPU quota, or
scheduling require recalibrating both software supply and device completion;
these values are not fixed thresholds across systems.

The probe-free complete-read results agree with this supply--demand decision:
P2 and P4 further accelerate submission but provide no end-to-end benefit. On
the measured large, concentrated read path, one executor supplies requests
more than an order of magnitude faster than the sustained device platform.
Fully utilizing an SSD therefore does not inherently require multiple
submission executors. Before adding cores, software should distinguish limits
from request supply, device completion, other host-side work, and request
organization.

\subsection{Linux Readahead and Concentrated Submission}
\label{sec:uc-readahead}

\subsubsection{Why Identical Reads Perform Differently}

We fix a single 128~MiB cold \texttt{pread} and change only the block-device
readahead window. With the default 128~KiB window, the median probe-free read
time is 184.41~ms; increasing the window to 2~MiB reduces it to 50.47~ms, a
$3.65\times$ difference. A seemingly simple configuration parameter can thus
substantially change the performance of the same synchronous read. To locate
the source of this difference, we first examine how Linux organizes and
triggers readahead requests.

On a cache miss, the measured Linux version determines a readahead window
from the current access state, issues I/O for the folios in that window, and
places a readahead marker within it. The reading thread obtains folios in file
order; when it reaches the marker, the asynchronous readahead path computes
and submits the next window. A later window cannot be triggered until the
thread advances to the marker in the newly issued window. Request supply is
therefore driven by read progress: the current window can overlap the next,
but arbitrarily many future windows do not reach the device at once. Treating
each window as a batch, this code structure forms a shallow pipeline with at
most two adjacent batches overlapping on the path studied here.

We verify this property with runtime events. For the $q$th observed readahead
batch, define its device in-flight interval as

\begin{equation}
I_q=[a_q,c_q],
\label{eq:uc2-batch-interval}
\end{equation}

where $a_q$ is the issue time of the batch's first device request and $c_q$
is the completion time of its last request. Sweeping all interval endpoints
gives the maximum number of simultaneously active batches,

\begin{equation}
d_{\max}=\max_t\#\{q:t\in I_q\}.
\label{eq:uc2-max-overlap}
\end{equation}

Across 27 complete traces with windows from 128~KiB to 64~MiB, every trace
satisfies $d_{\max}\leq2$; we never observe three active windows. The maximum
number of in-flight requests also grows with the window to nearly two
batches. The window directly determines batch shape: the 128~KiB setting
forms 1,025 single-request batches; 2~MiB forms 65 batches, usually with 16
requests each; and 64~MiB forms about three batches of roughly 512 requests
each. The source code explains the origin of the two-batch structure, and the
runtime timelines show that the measured execution follows it.

\subsubsection{Predicting Linux Completion Intervals from Device Batches}

To predict the completion time produced by this batched supply, we represent
a large read as a sequence of readahead windows. Each window corresponds to a
device-request batch whose duration can be predicted by the preceding device
model. If batch $q$ maps to $n_q$ equal-size device requests of size $S_q$,
its predicted duration from the first request issue to the last completion is

\begin{equation}
\widehat T_q=
F(n_q,S_q)+
\frac{(n_q-1)S_q}{B_{CQ}(n_q,S_q)}.
\label{eq:uc2-batch-prediction}
\end{equation}

Both $F$ and $B_{CQ}$ come from the independent device microbenchmark; the
target Linux traces are not used to fit them. Against the measured duration
of each batch, the prediction has a median absolute percentage error of
16.6\%.

The complete device-completion process can then be expressed as an interval
coverage problem. Let the union length of all readahead-batch intervals be

\begin{equation}
U=\left|\bigcup_q I_q\right|.
\label{eq:uc2-union}
\end{equation}

Because at most two batches are active at any instant on the measured path,
the observed intervals strictly satisfy

\begin{equation}
\frac{1}{2}\sum_qT_q\leq U\leq\sum_qT_q.
\label{eq:uc2-observed-bounds}
\end{equation}

Replacing each observed batch length with its independent device prediction
gives the cross-layer prediction range

\begin{equation}
\frac{1}{2}\sum_q\widehat T_q
\lesssim U\lesssim
\sum_q\widehat T_q.
\label{eq:uc2-predicted-bounds}
\end{equation}

\begin{figure*}[t]
  \centering
  \includegraphics[width=0.98\textwidth]{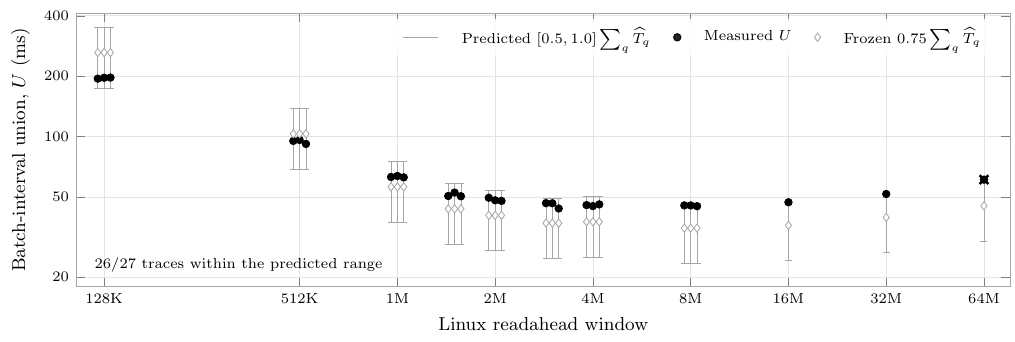}
  \caption{Predicting the union of Linux readahead-batch intervals. Vertical
  segments show the range derived from independently calibrated batch times
  and the measured two-batch overlap bound; filled circles show measured
  $U$, and open diamonds show the frozen $0.75\sum_q\widehat T_q$ estimate.}
  \label{fig:uc2-interval-prediction}
\end{figure*}

The lower bound represents two batches overlapping throughout, whereas the
upper bound represents no overlap. Figure~\ref{fig:uc2-interval-prediction}
shows all 27 traces. The range derived from independent device parameters and
the two-batch structure covers 26; in the sole miss, the measured $U$ at the
64~MiB boundary is only slightly above the predicted upper bound.

Before this experiment, we fixed an empirical coefficient $\alpha=0.75$ to
obtain the point estimate

\begin{equation}
\widehat U_{0.75}=0.75\sum_q\widehat T_q.
\label{eq:uc2-alpha-point}
\end{equation}

Its median absolute percentage error for $U$ is 17.4\%, making it a rough
single-value estimate. Applying the same estimate to the total span $H$ from
the first issue of the first batch to the last completion of the final batch
increases the median absolute percentage error to 29.2\%. $U$ counts only
periods in which at least one target batch remains in flight, whereas $H$
also includes gaps with no target request in flight; the fixed coverage
coefficient does not predict these gaps. The stable result is therefore the
range derived from the two-batch structure. The fixed 0.75 is not a
cross-device constant, and the predicted $U$ is neither thread blocking time
nor complete \texttt{pread} time.

We also vary the total read size as a holdout test. Without reselecting model
parameters or the coverage coefficient, all six 64~MiB and 256~MiB traces
retain a maximum active depth of two, and the precomputed two-batch ranges
cover every measured $U$. The fixed 0.75 remains only a rough point estimate.
Thus, the two-batch structure and interval range transfer across read sizes,
whereas a single empirical coverage coefficient is only auxiliary.

\subsubsection{Why the Default Small Window Is Slow}

The analysis exposes two simultaneous problems with the default 128~KiB
window. First, it limits how much work the device can see. Each 128~KiB window
usually maps to one 128~KiB request; even when two adjacent windows overlap,
the device normally sees only about two in-flight requests, far below the
independently measured sustained-completion platform. As the window grows,
each batch contains more requests and the shallow pipeline exposes more
effective in-flight work. Accumulated demand-folio wait falls from about
153~ms with 128~KiB to 17.6~ms with 2~MiB; the 32~MiB and 64~MiB diagnostic
traces show no such demand wait.

Second, every window introduces a new first-completion process, so $F$ can
repeatedly appear on the critical path. Linux can use the remainder of the
current batch after its first completion to cover part of the next batch's
startup wait, but that coverage does not always succeed. Let the progress
time from the current batch's first completion to the next batch's first
request issue be

\begin{equation}
\delta_q=a_{q+1}-c_{q,\mathrm{first}}.
\label{eq:uc2-delta}
\end{equation}

For equal-shape, ordered batches, a sufficient condition for the next batch's
first completion to arrive no later than the current batch's last completion
is

\begin{equation}
F(n,S)+\delta_q
\leq
\frac{(n-1)S}{B_{CQ}(n,S)}.
\label{eq:uc2-full-coverage}
\end{equation}

We compare this condition at every boundary with the direct timeline relation
$c_{q+1,\mathrm{first}}\leq c_{q,\mathrm{last}}$ and obtain 90.6\% agreement.
With the default window, $n=1$, so there is no post-first-completion interval
that can cover the next batch's startup, and the coverage margin degenerates
to $-F$. The independently calibrated $F(1,128\,\mathrm{KiB})$ is about
341.4~$\mu$s, and adjacent batches in the runtime traces only rarely achieve
completion overlap. The default window therefore both exposes too little
in-flight work to exploit aggregate device completion capability and
repeatedly exposes first-completion waits that the preceding batch tail does
not cover.

\subsubsection{Deriving a Readahead Window from the Model}

These two problems yield two complementary window-selection conditions.
Suppose one Linux readahead window normally contains $n$ device requests of
size $S$. Because at most two batches overlap, the device can see at most
about $2n$ requests. Let $Q_{\mathrm{sat}}(S)$ denote the smallest outstanding
depth at which an independent sustained-supply experiment reaches 90\% of the
device platform. The window should first satisfy

\begin{equation}
2n\geq Q_{\mathrm{sat}}(S).
\label{eq:uc2-platform-condition}
\end{equation}

The remaining completion interval after the current batch's first completion
should also cover the next batch's first-completion wait:

\begin{equation}
F(n,S)+\delta
\leq
\frac{(n-1)S}{B_{CQ}(n,S)}.
\label{eq:uc2-window-coverage-condition}
\end{equation}

The first condition prevents an undersized window from limiting device
completion capability; the second prevents each window from repeatedly
exposing startup wait. To obtain a candidate before observing the Linux
window sweep, we first set $\delta=0$ and use the independently calibrated
sustained-completion curve, $F$, and $B_{CQ}$ to find the smallest window
satisfying both conditions. This candidate need not globally minimize actual
Linux end-to-end time: the real $\delta$ includes folio inspection, copying,
scheduling, and submission preparation, while a large window may add I/O or
change the folio path.

For 128~KiB requests, the independent device experiment reaches its roughly
3.57~GB/s sustained platform at depth 32. Because Linux exposes at most two
adjacent windows, one window needs about 16 requests, or 2~MiB. At this point,
the batch's post-first-completion interval also exceeds the independently
calibrated first-completion wait. We therefore fixed 2~MiB as
\textsc{Linux-Model} before examining the Linux window sweep, and only then
performed the complete scan.

The sweep supports this choice. The 2~MiB window has a median read time of
50.47~ms, $3.65\times$ faster than the default; the empirical optimum at
64~MiB is 48.94~ms, only 3.13\% faster than the model point. Under the
predefined 3\% equivalence interval, the smallest platform point is 1.5~MiB,
at 50.28~ms. Larger windows do not improve performance monotonically: device
read amplification grows from $1.016\times$ at 2~MiB to about $1.20\times$ at
32/64~MiB, while a 128~MiB window reads exactly the requested bytes but
regresses to 67.79~ms because of the order-0 folio path analyzed earlier. The
model is valuable not because it guesses the fastest point in a discrete
sweep, but because it identifies, before the sweep, a configuration already
near the Linux performance platform while avoiding substantial overread and
a software-path reversal.

The model window is also reasonably robust to total read size. For 64~MiB and
256~MiB reads, 2~MiB outperforms the default by $3.44\times$ and
$3.57\times$, respectively. Yet for 64~MiB, 2~MiB is 10.5\% faster than an
8~MiB window, whereas for 256~MiB, 8~MiB is 5.1\% faster than 2~MiB. A fixed
configuration can provide a strong starting point without being optimal for
every read.

\begin{figure*}[t]
  \centering
  \includegraphics[width=0.96\textwidth]{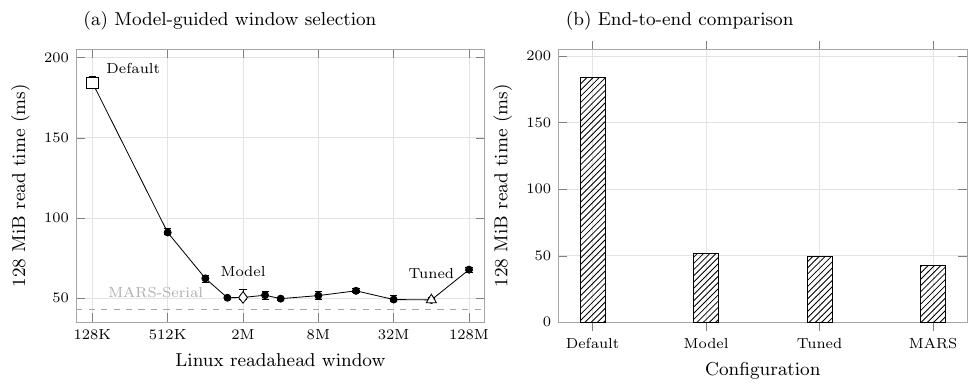}
  \caption{Model-guided readahead control and concentrated supply. (a) The
  complete Linux window sweep, with the model-selected point and device-byte
  amplification. (b) End-to-end comparison of default Linux,
  \textsc{Linux-Model}, empirically tuned Linux, and \textsc{MARS-Serial}.}
  \label{fig:uc2-window-control}
\end{figure*}

\subsubsection{From Window Tuning to Concentrated Submission}

Window tuning recovers most of the performance lost by default Linux, but it
has three boundaries. First, a suitable window depends jointly on the device
platform, request shape, and total read size; the holdout experiment already
shows that different read sizes can prefer different windows. Second, the
block-device readahead window is a system-wide setting: an application cannot
select an exact window for each \texttt{read}, and tenants in constrained
environments such as containers and serverless platforms often cannot control
the underlying block-device setting. Finally, a large window may fetch data
beyond the current demand. Whether that data is reused depends on the
workload; without reuse, it consumes device bandwidth and page-cache capacity.

These boundaries motivate a different request organization. When the system
already knows the complete target range of a read, it can supply that range
directly instead of relying on fixed windows triggered incrementally by read
progress. We implement this organization in the MARS experimental kernel
prototype and compare its single-submission-executor configuration,
\textsc{MARS-Serial}.

Concentrated supply also has a conditional theoretical advantage. Suppose it
and windowed supply ultimately process the same request set under fixed work
and resource constraints; concentrated supply makes every request available
no later, early availability does not reduce overall device service
capability, and it can still reproduce the scheduling order and deliberate
delays of the windowed scheme. Under these conditions, every feasible
windowed schedule is also feasible under concentrated supply. Its optimal
data-completion time therefore cannot be later.

In an ideal ordered pipeline with fixed $F$ and average completion interval
$p=S/B_{CQ}$, the extra stall introduced by window $q$ beyond the normal
completion interval is

\begin{equation}
E_q=[\delta_q+F-np]_+.
\label{eq:uc2-extra-stall}
\end{equation}

Thus,

\begin{equation}
T_{\mathrm{windowed}}
=T_{\mathrm{centralized}}+\sum_qE_q,
\qquad E_q\geq0.
\label{eq:uc2-windowed-vs-centralized}
\end{equation}

Here, $F+\delta\leq np$ merely prevents a stall longer than the normal
completion interval $p$. The earlier condition $F+\delta\leq(n-1)p$ is
stricter: it requires the next batch's first completion to arrive no later
than the current batch's last completion. Full coverage can reduce the
corresponding $E_q$ to zero and eliminate this disadvantage of windowed
supply, but delaying request availability cannot create an optimal device
completion earlier than concentrated supply. This claim concerns data
completion scheduling under the fixed-work and fixed-resource conditions
above; it does not imply that complete MARS \texttt{read} time unconditionally
beats Linux. Additional readahead, folio organization, copy overlap,
submission cost, and memory pressure can all affect a real system.

Finally, in the same experimental session, we compare default Linux,
\textsc{Linux-Model}, empirically tuned Linux, and \textsc{MARS-Serial}. Their
median times for a 128~MiB cold read are 183.97, 51.65, 49.77, and 42.73~ms,
respectively. The model window is $3.56\times$ faster than default Linux and
comes within 3.6\% of the empirical Linux optimum. Most of the gap between
default Linux and concentrated supply can therefore be removed by
understanding and correctly configuring the existing readahead mechanism.
\textsc{MARS-Serial} remains 17.3\% and 14.1\% faster than
\textsc{Linux-Model} and \textsc{Linux-Tuned}, respectively. These
configurations, however, do not perform strictly equal device work: the large
empirically tuned window incurs clear overread, whereas the actual read
volumes of \textsc{Linux-Model} and \textsc{MARS-Serial} are closer. The latter
pair therefore provides a closer equal-work system comparison; their
remaining gap can still include window boundaries, folio organization, copy
overlap, and a small amount of additional I/O.

This use case turns empirical parameter tuning into a testable reasoning
process using the device model and wait chain. Poor default synchronous-read
performance does not imply that the interface cannot exploit the SSD; it
arises from the shallow supply pipeline imposed by the readahead window.
Independently calibrated device-completion parameters predict the time scale
of Linux batches, and the two-batch structure then bounds the completion
interval of the whole read and yields a candidate window. A suitable Linux
configuration recovers most of the benefit of concentrated supply;
concentrated supply avoids fixed windows and incremental triggering and can
further improve completion time when the target work is precisely controlled.

\subsection{Dependent Reads and Data Layout}
\label{sec:uc-layout}

\begin{figure*}[!t]
  \centering
  \includegraphics[width=0.94\textwidth]{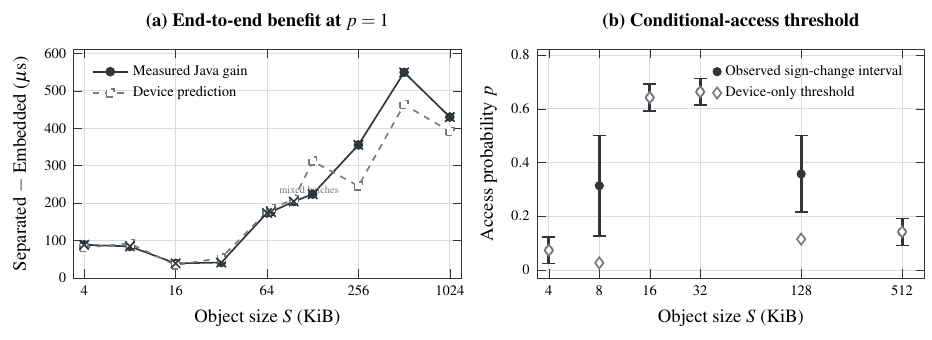}
  \caption{Predicting dependent-read layout decisions. (a) Benefits predicted
  from independent device parameters versus complete, probe-free Java-query
  measurements at $p=1$. Crosses mark the two mixed-request batches.
  (b) Model-predicted conditional-access thresholds and measured intervals
  between the last negative and first nonnegative tested benefits. The
  vertical ranges are sign-change intervals, not confidence intervals.}
  \label{fig:dependent-layout}
\end{figure*}

\subsubsection{From Two Dependent Reads to One Combined Read}

Arranging data according to dependencies between objects is a common data
placement strategy. For example, an application may first read an employee
record and then use the company identifier in that record to locate the
corresponding company record. If the two object types are stored separately,
the second read can be issued only after the first completes. If they are
stored contiguously, one combined read can retrieve both.

We evaluate this decision with a controlled Java Direct I/O microbenchmark.
\textsc{Separated} first reads an object $A$ of size $S$, obtains the location
of object $B$ from its contents, and then issues a second $S$-byte read.
\textsc{Embedded} stores the two objects contiguously and retrieves them with
one $2S$-byte read. The two layouts return the same logical result, and the
second read in \textsc{Separated} retains a true data dependency. Across 11
object sizes from 4~KiB to 1~MiB, \textsc{Embedded} always reduces complete
query time, but the benefit is not monotonic in object size. For example, the
mean savings are 173.83\,$\mu$s at 64~KiB per object and 224.06 and
549.25\,$\mu$s at 128 and 512~KiB, respectively.

The principal benefit does not come from eliminating one system call or one
traversal of the VFS path. At representative object sizes of 32, 128, and
512~KiB, the complete-query savings are 43.60, 222.73, and
546.62\,$\mu$s, whereas the differences in reading-thread CPU time are only
2.25, 1.06, and 1.37\,$\mu$s. The benefit therefore arises primarily from
changes in device waiting rather than from the host work saved by removing
one call. We next test whether reduced exposure to first-completion waiting
and changes in batch-drain time can explain the benefit quantitatively.

\subsubsection{Predicting Layout Benefits from Actual Request Mappings}

We first use the wait chain to recover the device-request streams produced by
the two application layouts and validate these mappings in independent
diagnostic runs. We then predict the application benefit using independent
device parameters. The mappings and all required device parameters are
sealed before inspecting the Java performance results, and the formal timing
runs carry no probes.

Let $A$ be the independently measured device time for reading one object and
$C$ the device-batch time produced by the combined read under its actual
request mapping. Because the two \textsc{Separated} reads are strictly
serial, their predicted layout benefit is

\begin{equation}
  \widehat{\Delta T}=2A-C.
  \label{eq:layout-benefit}
\end{equation}

The device model and wait-chain mapping yield three cases. First, when
$2S\le M$, where $M=128\,\mathrm{KiB}$ is the maximum request size on this
path, the combined read remains one request. Hence,

\begin{equation}
  A=F(1,S),\qquad C=F(1,2S),
\end{equation}

and

\begin{equation}
  \widehat{\Delta T}=2A-C=2F(1,S)-F(1,2S).
  \label{eq:layout-single}
\end{equation}

Second, when one object maps to $k$ equal requests of size $M$, the two
layouts produce batches of $k$ and $2k$ requests, respectively. Thus,

\begin{equation}
  A=T(k,M),\qquad C=T(2k,M),
\end{equation}

where

\begin{equation}
  T(n,M)=F(n,M)+\frac{(n-1)M}{B_{CQ}(n,M)}.
\end{equation}

Expanding Equation~\ref{eq:layout-benefit} gives

\begin{equation}
\begin{aligned}
  \widehat{\Delta T}
  ={}&2A-C \\
  ={}&2\left[F(k,M)+\frac{(k-1)M}{B_{CQ}(k,M)}\right] \\
     &-\left[F(2k,M)+\frac{(2k-1)M}{B_{CQ}(2k,M)}\right].
\end{aligned}
\label{eq:layout-homogeneous}
\end{equation}

Third, when $2S=kM+r$ and $0<r<M$, the combined read produces $k$ full
requests and one remainder request. We evaluate the full requests with the
equal-size batch model and approximate the remainder cost by bytes using the
batch's post-first-completion capability. For $k\ge2$,

\begin{equation}
\begin{aligned}
  C &\approx \widehat C
     \equiv T(k,M)+\frac{r}{B_{CQ}(k,M)} \\
    &=F(k,M)+\frac{(k-1)M+r}{B_{CQ}(k,M)}.
\end{aligned}
\label{eq:layout-remainder-cost}
\end{equation}

The corresponding benefit estimate is

\begin{equation}
  \widehat{\Delta T}
  =2A-\widehat C
  =2A-F(k,M)-\frac{(k-1)M+r}{B_{CQ}(k,M)}.
  \label{eq:layout-remainder-benefit}
\end{equation}

For the 68 and 96~KiB configurations, $2S=M+r$ and therefore $k=1$.
Because a single request has no post-first-completion interval, we approximate
the remainder cost using $B_{CQ}(2,M)$, the smallest multi-request batch.
Each \textsc{Separated} read is still one request, so $A=F(1,S)$.

Together, the three cases expose the actual tradeoff in combining dependent
reads. Combining the layouts reorganizes two serial device batches into one,
removing one first-completion wait from the critical path, but it also
requires draining a larger batch or one with a different request
composition. The benefit is therefore the difference between reducing one
exposed $F$ and changing the subsequent completion cost, rather than a fixed
saving from eliminating one system call.

Figure~\ref{fig:dependent-layout}(a) compares the independent device
prediction with complete Java-query measurements. For the nine object sizes
whose mappings are determined directly by the single-request or equal-size
batch equations, the model predicts every benefit direction correctly, with
absolute relative errors of 1.9--39.3\%. The predictions do not fit any
parameters to the application results; the residual errors include
differences in calibration conditions, fixed software costs, and unmodeled
effects in the application path. Despite amplitude errors at several points,
the device batches explain why the benefit varies nonmonotonically with
object size.

\subsubsection{Deriving a Layout Decision Rule}

Combined placement has a direct boundary: after reading the first object, an
application may not always need the second. Contiguous placement then trades
additional data transfer for the possible elimination of a serial wait, and
its benefit depends on the probability of accessing the second object. For
simplicity, we consider two equal-size objects and let $p$ be the probability
that the second object is accessed.

Let $A$ again denote the predicted device cost of reading one object and $C$
the predicted cost of the combined read. The expected times of the two
layouts are

\begin{equation}
  E[T_{\mathrm{sep}}]=(1+p)A,
  \qquad
  E[T_{\mathrm{emb}}]=C.
\end{equation}

The predicted benefit of \textsc{Embedded} over \textsc{Separated} is

\begin{equation}
  \widehat{\Delta}(p)=(1+p)A-C,
\end{equation}

which gives the layout-reversal threshold

\begin{equation}
  p^*=\frac{C}{A}-1.
  \label{eq:layout-threshold}
\end{equation}

When $p<p^*$, avoiding unnecessary reads favors \textsc{Separated}; when
$p>p^*$, eliminating the serial wait outweighs the extra transfer and favors
\textsc{Embedded}. This threshold follows from independent device parameters
and the actual request mapping, without first sweeping the application
results and fitting a crossing point.

We sweep $p$ at 4, 8, 16, 32, 128, and 512~KiB, using ten paired repetitions
per probability and sealing both the access sequence and prediction before
the experiment. All six sizes exhibit a layout reversal.
Figure~\ref{fig:dependent-layout}(b) compares predicted thresholds with the
measured intervals bounded by adjacent points with opposite benefit signs.
At 4, 16, 32, and 512~KiB, predicted thresholds of 0.074, 0.642, 0.663, and
0.142, respectively, fall inside the corresponding measured intervals. The
model therefore goes beyond explaining the benefit at $p=1$ and directly
provides useful layout-selection boundaries for these configurations.

The predictions at 8 and 128~KiB fall outside the measured reversal
intervals, showing that small prediction errors or layout-dependent fixed
costs can move the threshold. Consequently, $p^*$ is a starting point for an
advance decision rather than a precise constant across systems; when the
predicted benefit near the crossing is small, software controls or a
conservative interval are still necessary.

A natural engineering intuition is that contiguous placement should require
a higher access probability as objects grow. Compared with separate reads,
combining two objects prevents the second read from exposing a separate
first-completion wait. As the objects grow, however, unconditionally fetching
the second object transfers more data and progressively dilutes this
advantage. The application should therefore need greater confidence that the
second object will be accessed before contiguous placement becomes
worthwhile. Under this intuition, the layout-reversal threshold should
increase monotonically with object size.

The results contradict this monotonic expectation. The measured reversal
interval falls from 0.613--0.713 at 32~KiB to 0.215--0.500 at 128~KiB. The
wait chain attributes this decrease to a change in request mapping. Combining
two 32~KiB objects still produces one 64~KiB request, so the prediction uses
the first mapping case:

\[
  p^*_{32}
  =\frac{F(1,64\,\mathrm{KiB})}{F(1,32\,\mathrm{KiB})}-1
  =0.663.
\]

Combining two 128~KiB objects instead produces a batch of two 128~KiB
requests and therefore uses the second case:

\[
  p^*_{128}
  =\frac{T(2,128\,\mathrm{KiB})}{F(1,128\,\mathrm{KiB})}-1
  =0.115.
\]

The former mapping reads the additional data by enlarging one request. In the
latter, the additional request shares the batch's first-completion wait with
the first request and primarily pays the subsequent-completion cost. Growing
the objects thus changes both the byte count and how the device processes the
additional data, sharply reducing the marginal cost of contiguous placement.
Although the model underestimates the 128~KiB threshold, it correctly
predicts the direction of the decrease.

Changing the request mapping is not the only source of a lower threshold.
With the device-request size fixed at 128~KiB, independent device data still
predict that the threshold decreases from 0.374 for a 256~KiB object to 0.142
for a 512~KiB object. Larger objects reduce the relative benefit of sharing
one first-completion wait, while the larger batch simultaneously raises
$B_{CQ}$; these effects act in opposite directions. For the 256~KiB object,
expanding the read batch from two to four requests raises device time from
392.80 to 539.80~$\mu$s, an increase of 147.00~$\mu$s. For the 512~KiB
object, expanding it from four to eight requests adds only 76.66~$\mu$s. In
this region, the growth in $B_{CQ}$ offsets more of the additional-data cost,
so the threshold falls even as the object becomes larger. The measured
512~KiB reversal interval, 0.092--0.191, contains the prediction. Because the
application probability sweep did not include 256~KiB, this latter decrease
remains a mechanism-level prediction from the device model rather than a
complete end-to-end trend validation. Together, the two results show that
the layout threshold depends jointly on cross-layer request mapping and
batch-dependent completion capability, and need not increase monotonically
with object size.

This use case turns the empirical rule that related objects are usually
faster when placed together into a measurable cross-layer decision. The
application changes object placement, the software stack translates that
choice into one or more device requests, and first-completion waiting plus
subsequent completion determines the critical path. Combined with the actual
request mapping, independently calibrated $F$ and $B_{CQ}$ predict the
benefit direction for the measured regular batches and provide reversal
regions under conditional access. More importantly, the model and wait chain
expose two distinct reasons why the naive monotonic intuition fails:
discrete changes in request mapping alter how the device processes the
additional data, while batch-dependent $B_{CQ}$ changes the marginal
completion cost of fetching those data. Without either component, the
decrease in threshold as objects grow cannot be explained. Mixed batches and
layout-dependent fixed costs identify where the current model still requires
extension.

\subsection{F-Guided Polling}
\label{sec:uc-f-poll}

After submitting a request, software must still decide how to wait for its
completion. Normal interrupt-driven I/O (IRQ) allows the reading thread to
block, but observing a completion requires interrupt, wakeup, and scheduling
paths. Continuous polling (CP) checks the completion queue repeatedly and can
observe completions sooner, but it also consumes CPU throughout the
first-completion wait, when the device cannot yet return data. Linux Hybrid
Polling (LHP) lies between these alternatives. It divides requests into 16
statistical buckets by direction and size, with eight size classes for reads
and eight for writes. When an update is needed, LHP opens a 100-ms sampling
window and replaces a bucket with the mean completion time from the latest
nonempty window. A subsequent request first sleeps for approximately half of
its bucket value and then resumes polling. This mechanism requires no prior
device calibration, but it uses request size and direction as its entire
basis for distinguishing completion behavior. All requests in a bucket must
therefore share one prediction and one sleep duration.

\begin{figure*}[!t]
  \centering
  \includegraphics[width=0.96\textwidth]{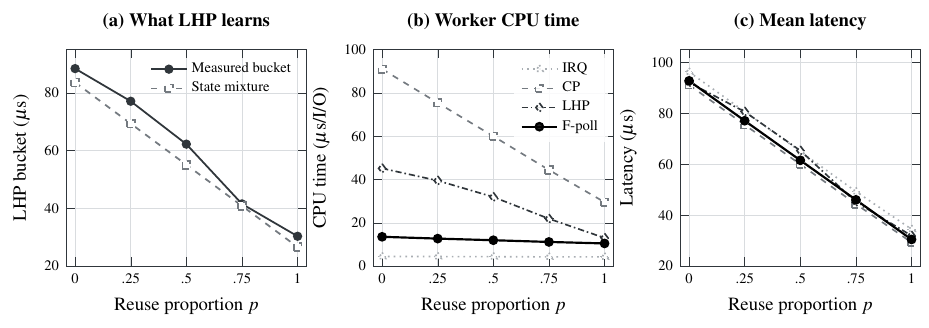}
  \caption{Device-state mixing and F-guided polling. (a) LHP bucket values
  predicted from independently measured state mixtures and their measured
  values. (b) Fio-worker CPU time under interrupt-driven I/O (IRQ), continuous
  polling (CP), Linux Hybrid Polling (LHP), and F-poll. (c) Mean application
  latency under the same policies. Error bars show 95\% confidence intervals
  over five diagnostic runs in (a) and ten probe-free runs in (b) and (c).}
  \label{fig:f-guided-polling}
\end{figure*}

\subsubsection{One Size Bucket Mixes Different Device States}

Equal-size requests need not have the same first-completion wait. On KI1, a
4-KiB request to a new address has a median SQ-to-CQ time of
84.882~$\mu$s, whereas immediately rereading the same complete 4-KiB range
reduces the median to 25.941~$\mu$s. Modern SSDs use controller DRAM or SRAM
to cache FTL address mappings and may also cache data; the NAND and controller
pipelines can retain short-term access state as well. Consequently, recent
address history can change the internal processing path and completion time
even when request sizes are identical. Linux still sees two equal-size block
requests and cannot infer the device's internal state from request size. We
do not assume which particular cache level produces KI1's fast completion;
we treat it as a host-observable device state.

We argue that an LHP bucket learns a weighted mixture of the completion times
of the device states represented in that bucket. Let $p$ denote the fraction
of measured requests that immediately and completely reuse an address. It is
a controlled software access state, not an unobservable SSD cache-hit rate.
Let $L_{\mathrm{cold}}$ and $L_{\mathrm{reuse}}$ be completion times measured
at the LHP statistical boundary in independent pure-state runs. A first-order
prediction of the bucket value is

\begin{equation}
\begin{aligned}
\widehat L_{\mathrm{LHP}}(p)
  &=(1-p)L_{\mathrm{cold}}+pL_{\mathrm{reuse}},\\
\widehat T_{\mathrm{sleep}}(p)
  &\approx \frac{\widehat L_{\mathrm{LHP}}(p)}{2}.
\end{aligned}
\label{eq:lhp-state-mixture}
\end{equation}

Independent calibration gives $L_{\mathrm{cold}}=83.62~\mu$s and
$L_{\mathrm{reuse}}=26.65~\mu$s. Before examining the target runs, these
values predict bucket values of 83.62, 69.37, 55.13, 40.89, and
26.65~$\mu$s for $p=0,0.25,0.5,0.75,$ and 1. The measured values in
Figure~\ref{fig:f-guided-polling}(a) are 88.42, 77.12, 62.27, 41.58, and
30.38~$\mu$s, giving a five-point mean absolute relative error of 9.10\% for
the linear prediction. Part of this error comes from LHP's own feedback:
resuming polling too late delays completion observation and further raises
subsequent bucket values. Adding this feedback in an offline calculation
reduces the error to 6.55\%. Even without that correction, the simple state
mixture captures the direction and principal time scale of the bucket's
change with workload composition. An LHP size bucket therefore does not learn
a unique device wait for that request size; it learns a mixture that changes
with the access composition.

\subsubsection{Scheduling Polling with State-Specific F}

Collapsing different device states into one bucket value underestimates the
first-completion wait of cold requests and overestimates that of fast reused
requests. These errors respectively increase busy-polling time and completion
observation latency. At $p=0.5$, for example, LHP sleeps for approximately
31.12~$\mu$s for both cold and reused requests. When polling resumes, 97.6\%
of cold requests are still incomplete and require another 52.1~$\mu$s of
busy polling on average. Conversely, 99.99\% of reused requests have already
completed at the first poll, indicating that software wakes too late for
them. One sleep value thus causes both excess polling for cold requests and
delayed completion observation for fast requests.

To address this problem, we implement F-poll. It independently calibrates
the first-completion wait $\widehat F(1,S,\mathrm{state})$ for request states
that software can distinguish and schedules polling to resume at an absolute
time relative to the SQ event:

\begin{equation}
t_{\mathrm{expire}}
=t_{\mathrm{SQ}}
+\widehat F(1,S,\mathrm{state})
-\widehat L_{\mathrm{wake}}-\delta.
\label{eq:f-poll-expiration}
\end{equation}

$\widehat L_{\mathrm{wake}}$ covers the wakeup cost from timer expiration
until the thread runs, and $\delta$ provides an additional early margin. The
absolute SQ time is essential: part of the first-completion wait has already
elapsed between submission and the waiting decision, so software must not
sleep for a full $F$ starting from the decision point. If the target time has
already passed, the thread polls immediately. Otherwise, each request sleeps
at most once and polls continuously after waking until completion.

Our prototype maintains access history by complete request interval
$(\mathrm{LBA},S)$. A range that has not been read successfully uses the cold
parameter; the same complete range uses the reuse parameter after a successful
read, and state changes only after successful completion. The experiment
constructs only two unambiguous states: new addresses and immediate complete
reuse. For 4-KiB requests, we freeze cold and reuse $F$ at 84.882 and
25.941~$\mu$s. The median, P95, and P99 steady-state wakeup costs are 4.723,
5.992, and 8.137~$\mu$s, respectively. The prototype uses P95 and adds a
2.146-$\mu$s early margin. The formal results include classification, lookup,
timer, and wakeup costs.

We compare IRQ, CP, restored LHP, and F-poll on the KI1 NVMe SSD using 4-KiB
fio \texttt{pvsync2} Direct I/O, one thread, and \texttt{iodepth=1}. Each
application I/O maps to one bio and one request, and
$p\in\{0,0.25,0.5,0.75,1\}$. Each configuration runs ten times. All four
policies use identical state sequences, and different repetitions and
policies use disjoint new physical-address ranges. Because current Linux has
removed the original LHP path, we restore
its bucket organization, 100-ms statistical window, and half-mean sleeping
semantics. We validate statistical updates, timed sleeping, split-request
fallback, and related paths in QEMU. QEMU is used only for functional
validation; all following performance results come from KI1.

Figure~\ref{fig:f-guided-polling}(b) shows that F-poll consumes
10.52--13.59~$\mu$s of fio-worker CPU time per I/O across the five state
mixtures, 64.4--85.1\% less than CP. Its mean latency in
Figure~\ref{fig:f-guided-polling}(c) is only 1.14--1.65~$\mu$s higher than
CP. Compared with IRQ, F-poll reduces mean latency by 3.26--4.02~$\mu$s. The
CPU metric in Figure~\ref{fig:f-guided-polling}(b) includes only the fio
worker and excludes interrupt processing on other CPUs.

State mixing most clearly exposes the limitation of LHP's size buckets. At
$p=0.25$ and $p=0.5$, F-poll reduces fio-worker CPU time relative to LHP by
67.7\% and 62.4\%, respectively, while lowering mean latency by 3.55 and
3.82~$\mu$s. If the latency reduction from LHP to continuous polling is used
as an empirical upper bound on what a waiting policy can recover in these
configurations, F-poll obtains 68.9\% and 70.8\% of that benefit while using
only 16.9\% and 20.0\% of CP's fio-worker CPU time. This agrees with the
mechanism diagnosis above: LHP can provide only one compromise sleep value
for both request states in a run, whereas F-poll places the polling-resumption
time for cold and fast requests shortly before their respective expected
completions. F-poll uses less fio-worker CPU than LHP at all five points.
These results show that an independently calibrated $F$ that preserves device
state can provide a more stable and interpretable CPU--latency tradeoff.

This result is currently limited to 4-KiB requests, $n=1$, and controlled
immediate complete reuse. Software access history is not a general oracle for
SSD cache state. This use case establishes that when software can identify a
state correlated with first-completion waiting, state-specific $F$ can
directly guide waiting control; it does not claim to solve online polling for
arbitrary requests and arbitrary device states.

\section{Discussion and Limitations}
\label{sec:discussion}

The device-completion model has three principal boundaries. First, its basic
object is a single request batch with explicit start and end boundaries. The
current model does not generally describe interactions among multiple batches
that share queue depth, device resources, and internal state. A later batch
may begin with a nonempty queue, and its first-completion wait may depend on
work from the preceding batch that has not yet drained.

Second, the current $T(n,S)$ assumes equal-size requests within a batch. A
batch containing multiple request sizes can experience different internal
splitting, scheduling, and parallel execution for each request, and its
completion order may change as well. The dependent-read use case estimates
batches with remainder requests using a byte-proportional approximation. It
provides useful estimates for the measured configurations, but is not a
general expression for arbitrary request sizes and completion orders.

Finally, $F$ and $B_{CQ}$ are external quantities calibrated for particular
devices, request sizes, batch sizes, access states, and supply conditions. We
use them for explanation, prediction, and decisions within the calibrated
domain. We do not extrapolate the finite measurement points to unmeasured
request shapes or device states, nor infer tail behavior such as P99 directly
from representative medians or means. When the device, request mapping, or
resource conditions change, these parameters must be recalibrated rather than
treated as fixed constants across systems.

Device-completion behavior alone also does not determine application
performance. End-to-end decisions must still account for host work, request
supply, cross-layer mappings, and their overlap with device waiting. Our
real-device evidence is currently concentrated on a limited number of SSDs
and controlled read workloads; validation across more devices, mixed
read/write workloads, and shared-device interference remains future work.

Overall, we position the model as a concise, calibratable explanation and
decision framework that can be connected across the software stack, rather
than a universal performance oracle for arbitrary I/O behavior. Its value is
not in retrospectively rewriting an observed time, but in retaining the key
temporal structure needed to distinguish request supply, repeatedly exposed
first-completion waits, data-dependency chains, and waiting policies. By
combining independent calibration, internal mechanisms, cross-layer mappings,
and actual interventions, and by testing gains, no gains, and gain reversals,
this work closes the loop from observation through modeling and understanding
to control within its studied domain. It thereby places empirical optimization
on testable performance reasoning.

\section{Related Work}
\label{sec:related}

\paragraph{I/O-stack measurement and performance models.}
Fast storage has made host software, interface choice, and device-internal
behavior increasingly important to end-to-end performance. Prior measurement
studies compare storage stacks including POSIX I/O, libaio, SPDK, and
\texttt{io\_uring}, and examine the application-level effects of many-core
processor caches, Linux I/O scheduling, and SSD performance
anomalies~\cite{RenTrivedi2023StorageStacks,LeeCluster2023ManycoreIO,
WhitakerHotStorage2023Schedulers,BorgeISPASS2020Anomalies}. They show that
software costs shift with devices and platforms, and that rated hardware
capability need not become application performance. We likewise begin with
end-to-end measurement, but separate host work, request supply, and device
waiting, and track how local time becomes exposed on the application critical
path rather than comparing only final throughput or latency across interfaces
and configurations.

Storage performance models provide another way to understand these effects.
Black-box models predict SSD latency, bandwidth, or IOPS from features such as
request size, queue depth, and access pattern; SSDcheck and LinnOS further
recover hidden device mechanisms or predict slow
requests~\cite{LiMASCOTS10BlackBoxSSD,KimTOS21SSDcheck,HaoOSDI20LinnOS}.
The Affine, PDAM, and PIO models incorporate initiation and transfer costs,
SSD parallelism, and read/write asymmetry into algorithm and data-structure
design~\cite{BenderTOPC21AffinePDAM,PaponDaMoN21PIO}. Shen et al. earlier
built a layered I/O model in which prefetching, scheduling, and the device each
transform the workload used for whole-system throughput
prediction~\cite{ShenFAST05ModelDebugging}. We inherit the idea of explaining
software performance through device calibration and inter-layer
transformations, but focus on the host-visible completion structure of a
finite request batch. $F$ preserves when the first completion becomes
visible, while $B_{CQ}$ describes how the remaining finite demand drains.
Both are calibrated across request and batch sizes and connected to software
blocking through the observed folio--bio--request mapping. We do not equate
them with fixed device latency and sustained bandwidth, nor aim to predict an
arbitrary single request or tail latency.

\paragraph{Host-side work and I/O-stack organization.}
One line of work adapts to fast devices by reducing page-cache management or
reorganizing execution. Async I/O Stack overlaps host tasks such as page
allocation, page-cache insertion, and DMA preparation with device processing;
HugeMap, StreamCache, and ScaleCache use larger management granularity,
dedicated page-cache pools, or scalable indexing structures to reduce
page-cache costs~\cite{LeeATC19AsyncIOStack,MalliotakisHugemap2021,
LiStreamCacheATC24,PhamEuroSys24ScaleCache}. Another line reduces data
copying through shared buffers, remapping, or specialized interfaces, as in
fbufs, IO-Lite, and Z-READ~\cite{DruschelPetersonFbufs1993,PaiIOLite2000,
ParkZREAD2019}. FlashShare, Async I/O Stack, and user-space or asynchronous
frameworks also shorten, bypass, or reorganize the traditional kernel path
for low-latency devices~\cite{ZhangOSDI18FlashShare,LeeATC19AsyncIOStack,
RenTrivedi2023StorageStacks}. In contrast, we first measure which costs
current Linux has already eliminated, amortized, or hidden: large folios
already reduce metadata work effectively on the appropriate path, and Linux
overlaps most cold-read copying with device waiting. We use MARS only as an
experimental kernel prototype for controlled interventions, evaluating more
aggressive out-of-order early copying, parallel execution of the remaining
copy, and parallel submission to synchronous backends. Our focus is when
these interventions shorten the critical path, and why the same parallel
submission does not benefit an SSD.

\paragraph{Request organization and data layout.}
Request merging, batching, and access reordering have long been used to
reduce calls and device accesses, enlarge lower-level I/O granularity, and
improve parallel I/O and scientific workflows~\cite{Thakur2002MPIIO,
Chowdhury2023ReqMerging,WangCherniack2006Permuting,Jensen2021FILCIO}.
In data systems, embedding frequently co-accessed objects in one record or
retaining references to avoid redundancy and update amplification is also a
standard layout tradeoff. Azure Cosmos DB's modeling guidance lists joint
querying, update frequency, object growth, and relationship shape among the
factors for choosing embedding or references~\cite{MicrosoftCosmosModeling,
MicrosoftCosmosPartitioning}. We do not propose a new merging or
denormalization technique. Instead, we ask whether device completion
structure can quantitatively support such decisions. Independently
calibrated $F$ and $B_{CQ}$ predict the benefit of merging two serial dependent
reads and, under conditional access, derive the reversal point between extra
bytes and eliminated dependent waiting. The case uses a controlled Java
Direct I/O path and does not cover update, consistency, or space-management
costs of a complete database.

\paragraph{I/O completion notification.}
Interrupts avoid continuous CPU occupation but incur wakeup and context-switch
costs; polling reduces completion-observation delay but may waste processor
time. Linux hybrid polling sleeps before polling, and subsequent work studies
the latency, CPU, and energy boundaries of interrupts, polling, and hybrid
polling, as well as polling paths in virtualized
environments~\cite{LeeJSA2022HybridPolling,HarrisHotStorage2022PollEnergy,
SeoHotStorage2024PolledIO}. Recent DPAS adapts sleep from recent I/Os and
switches among polling, interrupts, and adaptive sleeping to handle abrupt
latency changes, CPU contention, and erroneous sleep
feedback~\cite{SeoFAST26DPAS}. We do not propose a general completion method
covering these conditions, nor claim that F-poll replaces DPAS. Our case asks
a narrower modeling question: history aggregated only by request size can mix
hot and cold device states and feed delayed observations caused by sleeping
back into later estimates. We use independently calibrated, state-specific
$F$ to explain this bias and test whether scheduling polling and sleep from
first-completion wait can approach the latency benefit of pure polling while
reducing CPU use in the measured domain.

\section{Conclusion}
\label{sec:conclusion}

Using synchronous large-buffer Linux reads as an observation window, we
distinguished cumulative host work from critical-path exposure. Linux handles
metadata and cold-read data movement effectively through large folios and
I/O--copy overlap, but a local speedup becomes an end-to-end gain only when it
acts on the constrained stage. Our experimental kernel prototype further
demonstrated the value of more aggressive overlap and parallelism when these
mechanisms cease to apply.

We abstracted first-completion wait $F$ and subsequent completion capacity
$B_{CQ}$ from request-batch timelines and constructed a wait chain along the
actual folio--bio--request mapping. Evidence from a real device, simulation,
cross-layer prediction, and four classes of software decisions shows that the
framework connects device mechanisms to software waiting and supports
bottleneck identification, outcome prediction, and policy selection. Its value
lies in turning complex I/O behavior into calibratable, derivable, and
verifiable decision problems. This closes the loop from observation, modeling,
and understanding to control, providing a measurable and testable basis for
performance analysis and optimization choices in layered I/O stacks.

\appendix
\section{MARS Prototype Details}
\label{app:prototype-details}

\subsection{Out-of-Order Copy and Error Boundaries}
\label{app:early-copy-errors}

Out-of-order early copy changes when bytes are written to the user buffer, but
not how logical read progress is committed. Each early-copied folio records its
physical-copy progress without advancing the main iterator or file position.
The final Copy stage walks folios in file-offset order and advances the returned
prefix only across successful folios; data already copied are not copied again.
Thus, the return value and file position still describe a contiguous prefix.

The observable difference arises when lower-level I/O fails at an earlier file
offset after a later folio has already been copied. Linux's ordered copy path
normally leaves buffer regions beyond the returned prefix untouched, whereas
out-of-order early copy may already have modified such a region. MARS preserves
the returned prefix and file-position semantics but does not promise to retain
the previous contents of the user buffer beyond that prefix after such an
error. Out-of-order early copy can be disabled when sequential buffer-update
behavior is required.

\subsection{Opportunistic Worker Ownership and Recovery}
\label{app:worker-protocol}

The master partitions an operator into disjoint work items and retains one item
for local execution. A worker may execute an assigned item only after acquiring
its ownership atomically. If dispatch fails or the worker has not acquired the
item, the master cancels the assignment and executes that item itself. Once a
worker has acquired an item, the master does not execute it concurrently and
waits for the worker to publish either completion or a recoverable partial
state.

Before relinquishing an item, a worker publishes the operator-specific progress
needed for continuation, such as the copied-byte count, file position, folio
cursor, or first unsubmitted folio. State publication and observation are
ordered so that the master sees this recovery context before taking ownership.
The master then invokes the operator's restoration path and completes the
remainder. Consequently, workers can provide opportunistic acceleration and
yield their CPUs, while completion never depends on a worker remaining
available. All work items have converged to a completed state before the read
returns.

For copy, work items cover non-overlapping file and user-buffer ranges, and the
user mappings are made writable before worker execution. For submission, each
item owns complete readahead batches and records the first unfinished folio.
Workers use only backends that advertise the required nonblocking path; a
failure to make immediate progress transfers the remaining range to the master,
which continues through the ordinary sleepable path without resubmitting I/O
already issued successfully.

\subsection{Automatic Executor Selection}
\label{app:auto-selection}

Auto first derives an operator-specific upper bound $n_{\max}$ from the
remaining work, a minimum marginal-gain threshold $r$, the online CPU count,
and implementation limits. It then maps currently available workers to an
executor count. Let $N$ be the number of online CPUs, $q$ the number of workers
available at the decision point, $f=q/N$, and

\begin{equation}
  K=\min(n_{\max},q+1,N)
\end{equation}

be the largest feasible count including the master. If $q=0$, the master
executes alone. Otherwise, Auto approximates cumulative ideal gain as
$G(k)\propto 1-1/k$ and selects the smallest integer satisfying
$G(k)/G(K)\ge f^\alpha$, giving

\begin{equation}
  k=\left\lceil
  \frac{K}{K-(K-1)(q/N)^\alpha}
  \right\rceil.
  \label{eq:auto-executors}
\end{equation}

The parameter $\alpha$ controls how aggressively the policy consumes currently
available resources, while a larger $r$ lowers $n_{\max}$. Here $q$ is a
snapshot of MARS workers eligible to accept work, rather than an operating
system-wide idle-CPU fraction. The selected count is therefore an allocation
policy, not a prediction of the runtime-optimal count; dispatch acceptance and
the number of workers that ultimately complete their initial shares may be
smaller.

\section{Cross-Platform Replication on KI2}
\label{app:ki2-replication}

To test whether the main conclusions depend on KI1-specific device behavior,
we repeat the core evidence chain on the higher-specification KI2 instance. We
independently recalibrate every device parameter on KI2 rather than transferring
values from KI1. The replication covers the device-completion model, the
folio--bio--request mapping and cross-layer predictions, and the Linux
readahead-window and dependent-read-layout use cases, both of which include
complete cross-layer quantitative predictions. Table~\ref{tab:ki2-summary}
summarizes the results; the corresponding figures below present the replicated
results and key values.

\begin{table*}[t]
  \centering
  \small
  \caption{Summary of the KI2 cross-platform replication. Device parameters
  change substantially, while the completion structure, cross-layer
  propagation, and decision-level conclusions reproduce after recalibration.}
  \label{tab:ki2-summary}
  \begin{tabular}{p{0.30\textwidth}p{0.16\textwidth}p{0.16\textwidth}p{0.24\textwidth}}
    \toprule
    Evidence & KI1 & KI2 & Outcome \\
    \midrule
    $F(64,128\,\mathrm{KiB})$ & 575.7~$\mu$s & 192.2~$\mu$s & Lower on KI2 \\
    $B_{CQ}(64,128\,\mathrm{KiB})$ & 3.80~GB/s & 5.13~GB/s & Higher on KI2 \\
    Representative folio--bio--request mappings & Table~\ref{tab:cross-layer-mapping} & Same mappings & Reproduced \\
    Readahead-interval prediction coverage & 26/27 & 26/27 & Reproduced \\
    Model-selected Linux readahead window & 2~MiB & 4~MiB & Changed with device calibration \\
    Layout-benefit direction at $p=1$ & 11/11 & 11/11 & Reproduced \\
    Tested object sizes exhibiting a probability reversal & 6/6 & 9/9 & Reproduced \\
    \bottomrule
  \end{tabular}
\end{table*}

\subsection{Device-Completion Model}

\begin{figure*}[t]
  \centering
  \includegraphics[width=0.98\textwidth]{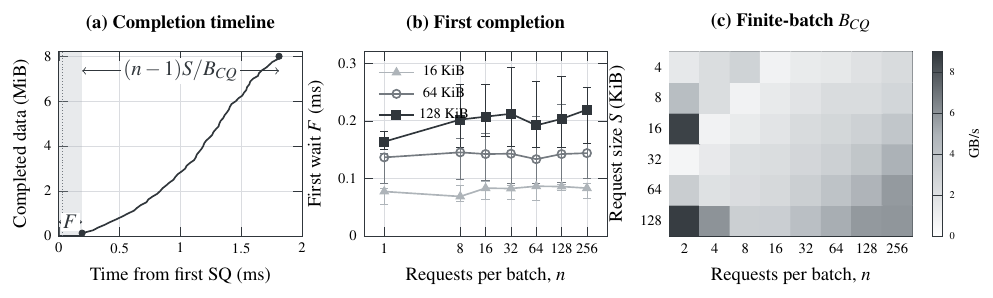}
  \caption{Device-completion model on KI2. (a) Cumulative completion timeline for a
  $128\,\mathrm{KiB}\times64$ batch; the shaded interval is $F$. (b) $F$ as
  request size and batch size vary. (c) The corresponding $B_{CQ}$ matrix.}
  \label{fig:ki2-device-model}
\end{figure*}

Figure~\ref{fig:ki2-device-model} uses the same configurations as the KI1
figure in the main text. KI2 retains the initial wait before the
first completion and the subsequent dense completion interval. For the
$128\,\mathrm{KiB}\times64$ batch, $F$ decreases from 575.7~$\mu$s on KI1 to
192.2~$\mu$s on KI2, while $B_{CQ}$ increases from 3.80 to 5.13~GB/s. The
higher-specification KI2 thus provides a substantially shorter
first-completion wait and greater subsequent completion capability. Despite
these different absolute values, the overall dependence of $F$ and $B_{CQ}$ on
request size and batch size remains. The completion structure therefore
reproduces across instances, but its parameters must be recalibrated on the
target device.

\subsection{Request Mapping and Cross-Layer Propagation}

Across 81 independent diagnostic runs, KI2 reproduces the representative
transformations in the main text: adjacent small bios are greedily merged up
to 128~KiB, larger bios are split into 128~KiB requests, and folio organizations
that form the same bio produce the same device-request stream.
Table~\ref{tab:ki2-mapping} lists representative mappings.

\begin{table*}[t]
  \centering
  \small
  \caption{Representative folio--bio--request mappings reproduced on KI2.}
  \label{tab:ki2-mapping}
  \begin{tabular}{p{0.48\textwidth}p{0.42\textwidth}}
    \toprule
    Folio or bio input & Device requests \\
    \midrule
    One 4, 16, 64, or 128~KiB bio & One request of the same size \\
    4, 8, 16, or 32 adjacent 4~KiB bios & One 16, 32, 64, or 128~KiB request \\
    One 256 or 512~KiB, or 1~MiB bio & 2, 4, or 8 requests of 128~KiB \\
    Six folio organizations totaling 128~KiB & One 128~KiB request \\
    64 4~KiB folios & One 256~KiB bio, split into two 128~KiB requests \\
    8 128~KiB folios & One 1~MiB bio, split into eight 128~KiB requests \\
    \bottomrule
  \end{tabular}
\end{table*}

Propagating independently calibrated KI2 batch times through these mappings
gives absolute errors of 0.97--9.98\% for the five normal \texttt{read\_pages}
points. In the strict-wave experiments, errors at both \texttt{submit\_bio}
and \texttt{read\_pages} remain at most 3.74\% for $q\geq4$, and every
configuration has the correct rank order. The single-request boundary
at $q=1$ has larger errors of 22.66--26.27\%. Overall, device-batch times
propagated through the observed mappings continue to predict upper-layer
waiting changes.

\begin{figure*}[t]
  \centering
  \includegraphics[width=0.92\textwidth]{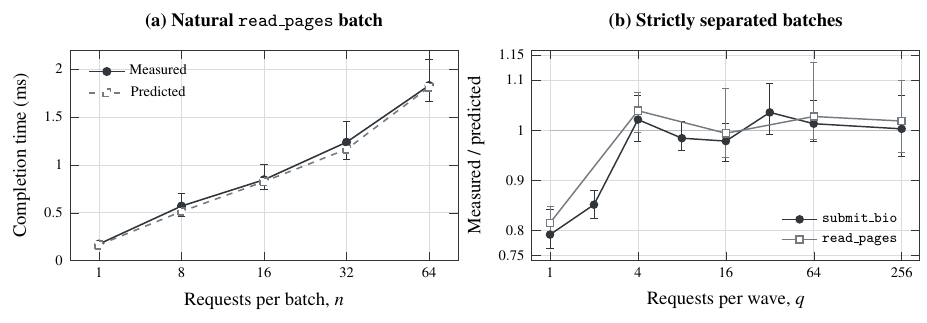}
  \caption{Cross-layer predictions on KI2. (a) Predicted and measured
  completion times for normal \texttt{read\_pages} batches. (b) Ratios of
  measured to predicted time for strict waves at \texttt{submit\_bio} and
  \texttt{read\_pages}.}
  \label{fig:ki2-wait-chain}
\end{figure*}

\subsection{Linux Readahead Windows}

All 27 KI2 readahead timelines retain a maximum of two simultaneously active
adjacent windows. The range derived from independent device-batch times covers
26 traces (96.30\%); the sole miss, at 64~MiB, is only about 0.33\% above the
predicted upper bound. The frozen $0.75\sum_q\widehat T_q$ point estimate has a
median absolute error of 7.02\% for the interval-union time.

Recalibration also changes the configuration decision. Before inspecting the
KI2 window scan, the model selects 4~MiB, compared with 2~MiB on KI1. The KI2
scan is fastest at 8~MiB. In the independent final comparison, the median times
for Default, Model, Tuned, and MARS are 84.88, 27.59, 25.37, and 23.00~ms. The
model therefore does not transfer an empirical KI1 setting: after recalibrating
the device, it selects a 4~MiB window that has already reached the performance
plateau.

\begin{figure*}[t]
  \centering
  \includegraphics[width=0.98\textwidth]{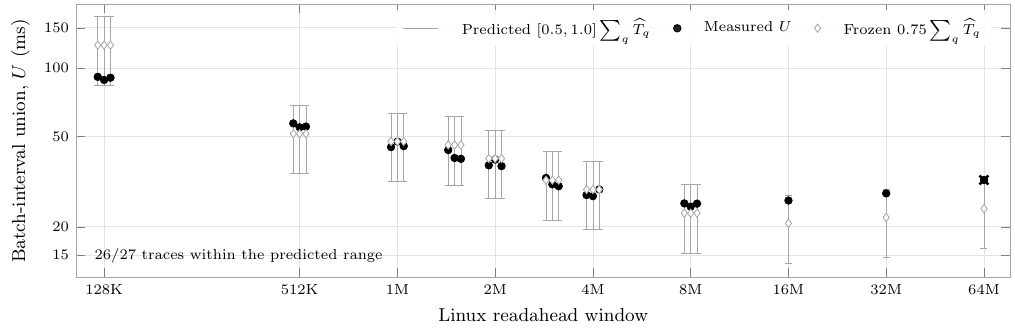}
  \caption{Readahead-interval prediction on KI2. Vertical segments show the
  range derived from independent batch times and the two-window overlap
  structure; filled circles show the measured interval union, and open
  diamonds show the frozen $0.75\sum_q\widehat T_q$ estimate.}
  \label{fig:ki2-readahead-interval}
\end{figure*}

\begin{figure*}[t]
  \centering
  \includegraphics[width=0.95\textwidth]{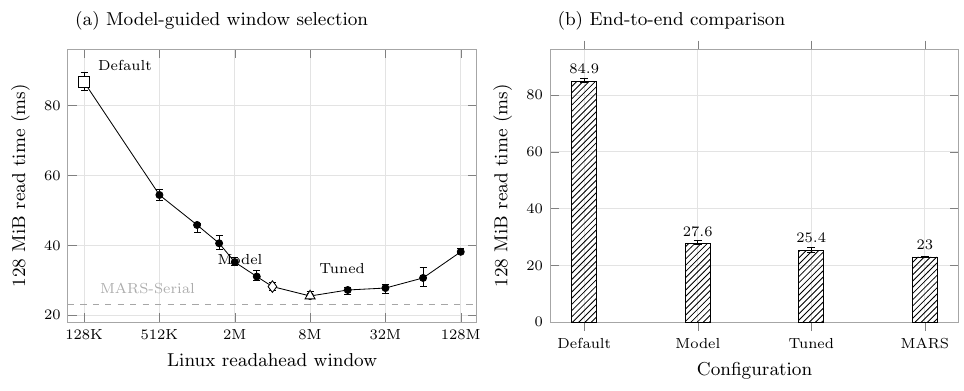}
  \caption{Readahead-window selection on KI2. (a) Linux window scan with the
  model-selected point frozen before the scan and the post-hoc tuned point.
  (b) Independent final comparison of Default, Model, Tuned, and MARS.}
  \label{fig:ki2-readahead-window}
\end{figure*}

\subsection{Dependent Reads and Data Layout}

The regular KI2 batches and the two heterogeneous supplements retain the
predicted benefit direction. We further define 61 frozen size--probability
configurations spanning nine object sizes from 4~KiB to 1~MiB and run 1,220
application trials, comprising ten paired comparisons per configuration. The
model predicts the benefit direction at 56/61 configurations, and every object size
exhibits a layout reversal. Predicted thresholds fall inside five of the nine
measured adjacent opposite-sign intervals; the remaining deviations are
concentrated near the crossing. The precise crossings and locations of the
nonmonotonic decreases differ from KI1, but the reversal and nonmonotonic shape
remain. What transfers is therefore the decision method based on the actual
request mapping, $F$, and batch-dependent $B_{CQ}$, rather than a fixed
probability threshold.

\begin{figure*}[t]
  \centering
  \includegraphics[width=0.94\textwidth]{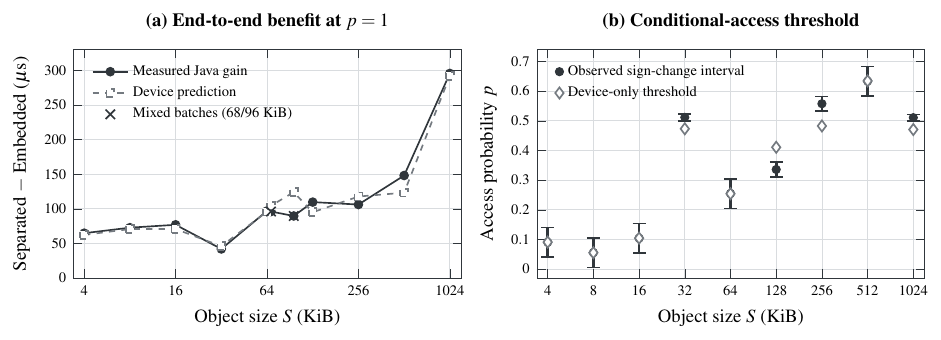}
  \caption{Dependent-read layout on KI2. (a) Merge benefits predicted from
  independent device measurements and measured by the complete Java query.
  (b) Model-predicted conditional-access thresholds and measured intervals
  between adjacent points with opposite benefit signs.}
  \label{fig:ki2-dependent-layout}
\end{figure*}

Overall, in the representative configuration, the higher-specification KI2
exhibits both a smaller $F$ and a higher $B_{CQ}$, and therefore substantially
greater device-completion capability than KI1. Despite changes in the absolute
parameters, model-selected window, and precise transition points, KI2
reproduces the request-batch completion structure, its propagation through the
observed mappings, and the main conclusions of both decision cases. The method
therefore requires calibration on the target platform, but does not depend on
KI1-specific parameter values or tuning rules.

\clearpage

\bibliographystyle{plain}
\bibliography{references}
\end{document}